\documentclass[11pt]{article}
\usepackage{mathrsfs}
\usepackage{amsfonts}
\usepackage{amsmath}
\usepackage{amssymb,amsthm}
\usepackage{hyperref,color}
\usepackage{exscale,relsize}

\numberwithin{equation}{section}
\allowdisplaybreaks[4]

\author{ Yu Hou   \and  Engui Fan\footnote{Corresponding
author and  e-mail address:
      faneg@fudan.edu.cn}}
\date{   \small{ School of Mathematical Sciences,  Shanghai Center for Mathematical Sciences \\
 and Key Laboratory of Mathematics for Nonlinear Science, \\ Fudan
University, Shanghai 200433, P.R. China}}
\title{\bf \Large{Algebro-geometric solutions for the two-component Camassa-Holm Dym hierarchy} }
\begin{document}
\maketitle

\begin{abstract}
This paper is dedicated to provide theta function representations of
algebro-geometric solutions and related crucial quantities for the
two-component Camassa-Holm Dym (CHD2) hierarchy.
Our main tools include the polynomial
recursive formalism, the hyperelliptic
curve with finite number of genus, the Baker-Akhiezer functions, the
meromorphic function, the Dubrovin-type equations for auxiliary
divisors, and the associated trace formulas. With the help of these
tools, the explicit representations of the algebro-geometric solutions  are
obtained for the entire CHD2 hierarchy.\\ 
{\bf Key words:}  two-component Camassa-Holm Dym,  hyperelliptic
curve,  algebro-geometric solutions    
\end{abstract}

\section{Introduction}
In this paper we consider the following integrable two-component Camassa-Holm Dym (CHD2)
system:
 \begin{equation}\label{1.1}
  \left\{
    \begin{array}{ll}
      \rho_t+\left(\displaystyle \frac{m}{\rho^2}\right)_x=0,  \\
      m_t-\left(\displaystyle (1-\partial_x^2)\frac{1}{\rho}\right)_x=0,
    \end{array}
  \right.
 \end{equation}
where $m=u-u_{xx}$, which was recently introduced by Holm and Ivanov in \cite{Holm001}.
 This coupled nonlinear system is
at the position in the two component Camassa-Holm  hierarchy
\cite{chen, Constantin01, Escher, chuanx, Gui, henrydj, Ivanov, Johnson, song}
 that corresponds to
the modified Dym equation, first introduced as a tri-Hamiltonian system in \cite{Olver}.
The CHD2 equation  combines to produce the nonlinear wave equation
  \begin{equation}\label{1.2}
   m_{tt}=(1-\partial_x^2)\left(
   \partial_x \frac{1}{\rho^2}
   \partial_x \frac{1}{\rho^2} \right)m.
  \end{equation}
Linearizing this equation around $m=0$ and $\rho=1$
yields the dispersion relation for a plane wave $\mathrm{exp} (i (\kappa x-wt))$
with wave number $\kappa$ and frequency $w$ as
  \begin{equation*}
   w^2(\kappa )=(1+\kappa^2)\kappa^2.
  \end{equation*}
Accordingly, the phase speed of the linearized plane waves is
$w/\kappa=\sqrt{1+\kappa^2}$,
so the higher wave numbers travel faster \cite{Holm001}.
This type of dispersion relation is the same as for time-dependent
Euler-Bernoulli theory for an elastic beam with
both bending and vibration response \cite{Holm001,reddy}.

The CHD2 equation is also a completely integrable system with a bi-Hamiltonian structure,
and hence, it possesses a Lax pair and conservation laws \cite{Holm001}.
Recently, soliton solutions and travelling wave solutions of
the system (\ref{1.1}) were investigated in \cite{Holm001}.
However, within the knowledge of the authors, the
algebro-geometric solutions of the entire CHD2 hierarchy are not studied yet.

 The principal subject of this paper concerns algebro-geometric quasi-periodic solutions
 of the whole CHD2 hierarchy, of which (\ref{1.1}) is just the first of infinitely many members.
 Algebro-geometric solution, as an important feature of integrable system,
 is a kind of explicit solution closely related to the inverse spectral
 theory \cite{4}-\cite{7}, \cite{9}-\cite{11}. In a degenerated case of the
 algebro-geometric solution, the multi-soliton solution and periodic solution
 in elliptic function type may be obtained \cite{7, 33, 8}.
 A systematic approach, proposed by Gesztesy and Holden to construct
 algebro-geometric solutions for integrable equations,   has been extended to
 the whole (1+1) dimensional integrable hierarchy, such as the AKNS hierarchy,
 the Camassa-Holm (CH) hierarchy etc. \cite{13}-\cite{12}.
 Recently, we investigated algebro-geometric solutions for the
 modified CH hierarchy and the Degasperis-Procesi hierarchy \cite{18, 17}.

 The outline of the present paper is as follows.

 In section 2, based
 on the polynomial recursion formalism, we derive the CHD2 hierarchy,
 associated with the $2 \times 2$ spectral problem. A hyperelliptic curve
 $\mathcal{K}_{n}$ of arithmetic genus $n$ is introduced with
 the help of the characteristic polynomial of Lax matrix $V_n$ for the
 stationary CHD2 hierarchy.

 In Section 3,  we decompose the stationary
 CHD2 equations into a system of Dubrovin-type
 equations. Moreover, we obtain the stationary trace formulas
 for the CHD2 hierarchy.

 In Section 4, we present the first set of our results, the explicit
 theta function representations of the Baker-Akhiezer function, the meromorphic function,
 and in particular, that of the potentials $u, \rho$
 for the entire stationary CHD2 hierarchy.

 In Sections 5 and 6, we extend the analyses of Sections 3 and 4, respectively,
 to the time-dependent case. Each equation in the CHD2
 hierarchy is permitted to evolve in terms of an independent time
 parameter $t_r$. As  initial data, we use a stationary solution of
 the $n$th equation and then construct a time-dependent solution of
 the $r$th equation of the CHD2 hierarchy.
 The Baker-Akhiezer function,  the analogs
 of the Dubrovin-type equations, the trace formulas, and the theta function
 representations in Section 4 are all extended to the time-dependent
 case.

 Finally, it should perhaps be noted that the system (\ref{1.1}) was obtained by using the
 condition $u \rho =-2$ in \cite{Holm001}.  At this point, we will occasionally use this relation in the
 following sections.

\section{The CHD2 hierarchy}
 In this section, we provide the construction of  CHD2 hierarchy and derive the
 corresponding
 sequence of zero-curvature pairs using a polynomial recursion formalism.
 Moreover, we introduce the underlying hyperelliptic curve in connection with
 the stationary CHD2 hierarchy.

 Throughout this section, we make the following hypothesis.

\newtheorem{hyp1}{Hypothesis}[section]
 \begin{hyp1}
      In the stationary case, we assume that
  \begin{equation}\label{2.1}
    \begin{split}
    & u, \rho \in C^\infty(\mathbb{R}), \quad u(x) \neq 0, \rho(x) \neq 0, \quad
   ~\partial_x^k u, \partial_x^k \rho  \in L^\infty (\mathbb{R}), \quad k\in \mathbb{N}_0.
    \end{split}
  \end{equation}
    In the time-dependent case, we suppose
  \begin{equation}
    \begin{split}\label{2.2}
    & u(\cdot,t), \rho(\cdot,t) \in C^\infty (\mathbb{R}),\ \
     \partial_x^ku(\cdot,t), \partial_x^k \rho(\cdot,t) \in L^\infty
    (\mathbb{R}), ~~ k\in \mathbb{N}_0,~ t\in \mathbb{R},\\
    & u(x,\cdot), u_{xx}(x,\cdot), \rho(x,\cdot), \rho_x(x, \cdot) \in C^1(\mathbb{R}),
       \quad x\in \mathbb{R},\\
    & u(x,t)\neq 0, \rho(x,t) \neq 0, \quad (x,t)\in \mathbb{R}^2.
  \end{split}
  \end{equation}
\end{hyp1}

We first introduce the basic polynomial recursion formalism. Define
 $\{f_{ l}\}_{l\in\mathbb{N}_{0}}$,
 $\{g_{ l}\}_{l\in\mathbb{N}_{0}}$, and
 $\{h_{ l}\}_{l\in\mathbb{N}_{0}}$
recursively by
 \begin{equation}\label{2.3}
      \begin{split}
        & f_0=-u, \\
       & f_{l}=\mathcal{G}\left(-f_{l-2,xxx}+4(u-u_{xx})f_{l-1,x}+f_{l-2,x}+2(u_x-u_{xxx})f_{l-1}\right),
        \quad l \in \mathbb{N},\\
      & g_l=\frac{1}{2}f_{l,x},\quad l \in \mathbb{N}_0,\\
      & h_l=-g_{l-2,x}-\rho^2f_{l}+(u-u_{xx})f_{l-1}+\frac{1}{4}f_{l-2}, \quad l \in \mathbb{N}_0,\\
      \end{split}
 \end{equation}
 where $\mathcal{G}$ is given by
    \begin{equation}\label{2.4}
     \begin{split}
      & \mathcal{G}: L^\infty(\mathbb{R}) \rightarrow
        L^\infty(\mathbb{R}), \\
     & (\mathcal{G}v)(x)=\frac{1}{4}\rho(x)^{-1}
     \int_{-\infty}^x    \rho(y)^{-1} v(y)~dy, \quad x\in\mathbb{R},~
        v\in L^\infty(\mathbb{R}).
      \end{split}
    \end{equation}
One observes that
    \begin{equation}\label{2.5}
        \mathcal{G}=\left( 2\partial_x \rho^2 +2 \rho^2 \partial_x \right)^{-1}.
    \end{equation}
Explicitly, one computes
   \begin{equation}\label{2.6}
     \begin{split}
      & f_0=-u,\\
      & f_1=\mathcal{G}(-4u_x(u-u_{xx})-2u(u_x-u_{xxx}))+c_1(-u),\\
      & g_0=-\frac{1}{2}u_x,\\
      & g_1=\frac{1}{2} \mathcal{G}(-4u_{xx}(u-u_{xx})-6u_x(u_x-u_{xxx})-2u(u_{xx}-u_{xxxx}))+c_1(-\frac{1}{2}u_x),\\
      & h_0=u\rho^2,\\
      & h_1=-u(u-u_{xx})-\rho^2f_1,~\mathrm{etc}.,\\
     \end{split}
   \end{equation}
where $\{c_l\}_{l\in\mathbb{N}}\subset\mathbb{C}$ are integration
constants.

Next, it is convenient to introduce the corresponding homogeneous coefficients
$\hat{f}_{l}, \hat{g}_{l},$ and  $\hat{h}_{l},$ defined by the vanishing of the
integration constants $c_k,k=1,\ldots,l,$
 \begin{equation}\label{2.8}
       \begin{split}
        & \hat{f}_0=f_0=-u, \quad \hat{f}_{l}=f_{l}|_{c_k=0,~k=1,\ldots,l},
         \quad l \in \mathbb{N},
        \quad \\
        & \hat{g}_0=g_0=-\frac{1}{2}u_x, \quad
        \hat{g}_{l}=g_{l}|_{c_k=0,~k=1,\ldots,l},
        \quad  l \in \mathbb{N},\\
        &   \hat{h}_0=h_0=u\rho^2, \quad
           \quad \hat{h}_{l}=h_{l}|_{c_k=0,~k=1,\ldots,l},
           \quad l \in \mathbb{N}.
       \end{split}
    \end{equation}
Hence,
    \begin{equation}\label{2.9}
      f_{l}=\sum_{k=0}^{l}c_{l-k}\hat{f}_{k}, \quad
      g_{l}=\sum_{k=0}^{l}c_{l-k}\hat{g}_{k}, \quad
      h_{l}=\sum_{k=0}^{l}c_{l-k}\hat{h}_{k}, \quad
       l\in \mathbb{N}_0,
    \end{equation}
defining
    \begin{equation}\label{2.10}
        c_0=1.
    \end{equation}
Now, given Hypothesis 2.1,
one introduces the following $2\times 2$ matrix $U$ by
     \begin{equation}\label{2.11}
       \psi_x=U(z,x)\psi=
       \left(
         \begin{array}{cc}
           0 & 1 \\
           -z^{2}\rho^2+z(u-u_{xx})+\frac{1}{4} & 0 \\
         \end{array}
       \right)
       \psi,
     \end{equation}
and for each $n\in\mathbb{N}_0,$ the following $2\times 2$
matrix $V_n$ by
\begin{equation} \label{2.12}
       \psi_{t_n}=V_n(z)\psi,
    \end{equation}
  with
    \begin{equation}\label{2.13}
      V_n(z)=
      \left(
        \begin{array}{cc}
          -zG_n(z) & zF_{n}(z) \\
          zH_n(z) & zG_n(z) \\
        \end{array}
      \right),
    \quad z \in \mathbb{C} \setminus \{0\},
    \quad n\in\mathbb{N}_0,
    \end{equation}
assuming $F_{n}$, $G_n$, and $H_n$ to be polynomials\footnote{$F_{n}, G_n, H_n$ are
polynomials of degree $n,n,n+2$, respectively.}
with respect to $z$ and $C^{\infty}$ in $x$. The compatibility
condition of linear system (\ref{2.11}) and (\ref{2.12}) yields the
stationary zero-curvature equation
\begin{equation}\label{2.14}
   -V_{n,x}+[U,V_n]=0,
\end{equation}
which is equivalent to
   \begin{align}
      & F_{n,x}= 2G_n, \label{2.15a}\\
      & H_{n,x}= -2\left(-z^2\rho^2+z(u-u_{xx})+\frac{1}{4}\right)G_n, \label{2.15b}\\
      & G_{n,x}=-H_n+\left(-z^2\rho^2+z(u-u_{xx})+\frac{1}{4}\right)F_{n}.\label{2.15c}
    \end{align}
From (\ref{2.15a})-(\ref{2.15c}), one infers that
   \begin{equation}\label{2.18}
     \frac{d}{dx} \mathrm{det} (V_n(z,x))=
     -z^2 \frac{d}{dx} \Big(
     G_n(z,x)^2+F_{n}(z,x)H_n(z,x)
     \Big)=0,
   \end{equation}
and hence
   \begin{equation}\label{2.19}
    G_n(z,x)^2+F_{n}(z,x)H_n(z,x)=R_{2n+2}(z),
   \end{equation}
where the polynomial $R_{2n+2}$ of degree $2n+2$ is $x$-independent.
In another way, one can write
$R_{2n+2}$ as
\begin{equation}\label{2.20}
   R_{2n+2}(z)=-4\prod_{m=0}^{2n+1}(z-E_m),\quad
   \{E_m\}_{m=0,\ldots,2n+1}\in\mathbb{C}.
   \end{equation}

Next, we compute
the characteristic polynomial $\mathrm{det}(yI-z^{-1} V_n)$ of Lax matrix
$z^{-1} V_n$,
    \begin{align}\label{2.26}
      \mathrm{det}(yI-z^{-1} V_n)&=
        y^2-G_n(z)^2-F_{n}(z)H_n(z)
         \nonumber \\
      &= y^2-R_{2n+2}(z)=0,
    \end{align}
and then introduce the (possibly singular) hyperelliptic curve
$\mathcal{K}_n$ of arithmetic genus $n$ defined by
\begin{equation}\label{2.27}
     \mathcal {K}_n:\mathcal {F}_n(z,y)=y^2-R_{2n+2}(z)=0.
   \end{equation}
 In the following, we will occasionally impose further constraints on
 the zeros $E_m$ of $R_{2n+2}$ introduced in (\ref{2.20}) and assume
 that
 \begin{equation}\label{2.26a}
  E_m\in\mathbb{C}, \quad E_{m}\neq E_{m^\prime},\quad
 \forall m\neq m^\prime,\quad  m,m^\prime=0,\ldots,2n+1.
 \end{equation}

The stationary zero-curvature equation
(\ref{2.14}) implies polynomial recursion relations (\ref{2.3}).
Introducing the following polynomials $F_{n}(z), G_n(z)$, and
$H_n(z)$ with respect to the spectral parameter $z$,
   \begin{align}
    & F_{n}(z)=\sum_{l=0}^{n} f_{l} z^{n-l},\label{2.28}
         \\
    &
       G_n(z)=\sum_{l=0}^{n} g_{l} z^{n-l}, \label{2.29}
         \\
    &
      H_n(z)=\sum_{l=0}^{n+2} h_{l} z^{n+2-l}. \label{2.30}
   \end{align}
Inserting (\ref{2.28})-(\ref{2.30}) into (\ref{2.15a})-(\ref{2.15c}) then yields
the recursion relations (\ref{2.3}) for $f_{l},$ $l=0,\ldots, n,$
and $g_{l}$, $l=0,\ldots, n.$  For fixed $n \in \mathbb{N}_0$,
we obtain the recursion relations for $h_{l}, $ $l=0, \ldots,
n$ in (\ref{2.3}) and
  \begin{equation}\label{2.31}
     \begin{split}
        & h_{n+1}=-\frac{1}{2}f_{n-1,xx}+(u-u_{xx})f_{n}+\frac{1}{4}f_{n-1}, \\
        &   h_{n+2}=-\frac{1}{2}f_{n,xx}+\frac{1}{4}f_n.
     \end{split}
    \end{equation}
Moreover, from (\ref{2.15b}), one infers that
   \begin{equation}\label{2.32}
     \begin{split}
    &-h_{n+1,x}-(u-u_{xx})f_{n,x}-\frac{1}{4}f_{n-1,x}=0, \quad n\in \mathbb{N}_0,\\
    &   -h_{n+2,x}-\frac{1}{4}f_{n,x}=0, \quad n\in \mathbb{N}_0.\\
     \end{split}
   \end{equation}
Then using  (\ref{2.31}) and (\ref{2.32})
permits one to write the stationary  CHD2 hierarchy as
 \begin{equation}\label{2.34}
        \textrm{s-CHD2}_n(u,\rho)=
        \left(
          \begin{array}{c}
            -2(u-u_{xx})f_{n,x}-(u_x-u_{xxx})f_n-\frac{1}{2}f_{n-1,x}+\frac{1}{2}f_{n-1,xxx} \\
            -\frac{1}{2}f_{n,x}+\frac{1}{2}f_{n,xxx} \\
          \end{array}
        \right)=0,
        \quad n\in \mathbb{N}_0.
      \end{equation}
We record the first equation explicitly,
    \begin{equation}\label{2.35}
      \begin{split}
       & \textrm{s-CHD2}_0(u,\rho)=
        \left(
         \begin{array}{c}
           2u_x(u-u_{xx})+u(u_x-u_{xxx}) \\
           \frac{1}{2}u_x-\frac{1}{2}u_{xxx} \\
         \end{array}
       \right)=0.
       \end{split}
    \end{equation}

By definition, the set of solutions of (\ref{2.34}) represents the
class of algebro-geometric CHD2 solutions, with $n$ ranging in
$\mathbb{N}_0$ and $c_l$ in $\mathbb{C},~l\in\mathbb{N}$. We call
the stationary algebro-geometric CHD2 solutions $u,$ $\rho$ as CHD2 potentials
at times.

\newtheorem{rem2.2}[hyp1]{Remark}
  \begin{rem2.2}
    Here, we emphasize that if $u$, $\rho$ satisfy one of the stationary CHD$2$ equations
    in
    $(\ref{2.34})$ for a particular value of $n$, then they satisfy infinitely many such equations
    of order higher than $n$ for certain choices of integration
    constants $c_l$. This is a common characteristic of the general
    integrable soliton equations such as the KdV, AKNS, and CH
    hierarchies \cite{15}.
  \end{rem2.2}

Next, we introduce the corresponding homogeneous polynomials
$\widehat{F}_{l}, \widehat{G}_{l}, \widehat{H}_{l}$ by
    \begin{eqnarray}
    &&
      \widehat{F}_{l}(z)=F_{l}(z)|_{c_k=0,~k=1,\dots,l}
      =\sum_{k=0}^{l} \hat{f}_{k} z^{l-k},
      ~~ l=0,\ldots,n, \label{2.36a1}\\
    &&
      \widehat{G}_{l}(z)=G_{l}(z)|_{c_k=0,~k=1,\dots,l}
      =\sum_{k=0}^l \hat{g}_{k} z^{l-k},
      ~~ l=0,\ldots,n,\\
    &&
      \widehat{H}_{l}(z)=H_{l}(z)|_{c_k=0,~k=1,\dots,l}
      =\sum_{k=0}^l \hat{h}_{k} z^{l-k},
      ~~ l=0,\ldots,n,\\
    &&
      \widehat{H}_{n+1}(z)= -\hat{g}_{n-1,x}+(u-u_{xx})\hat{f}_{n}+\frac{1}{4}\hat{f}_{n-1}+
      \sum_{k=0}^{n} \hat{h}_{k} z^{n+1-k}, \\
    &&
       \widehat{H}_{n+2}(z)= -\hat{g}_{n,x}+\frac{1}{4}\hat{f}_{n}+ z\widehat{H}_{n+1}(z).\label{2.36a2}
          \end{eqnarray}

In accordance with our notation introduced in (\ref{2.8}) and (\ref{2.36a1})-(\ref{2.36a2}),
the corresponding homogeneous stationary CHD2
 equations are then defined by
      \begin{equation}\label{2.40}
        \textrm{s-}\widehat{\mathrm{CHD2}}_n(u,\rho)=
        \textrm{s-CHD2}_n(u,\rho)|_{c_l=0,~l=1,\dots,n}=0,
        \quad n\in\mathbb{N}_0.
      \end{equation}

At the end of this section, we turn to the time-dependent CHD2 hierarchy. In this case, $u$, $\rho$
are considered as  functions of both space and time. We introduce
a deformation parameter $t_n \in \mathbb{R}$ in $u$ and $\rho$, replacing
$u(x),\rho(x)$ by $u(x,t_n), \rho(x,t_n)$, for each equation in the hierarchy. In
addition, the definitions (\ref{2.11}), (\ref{2.13}),
and (\ref{2.28})-(\ref{2.30}) of $U,$ $V_n$ and $F_{n}, G_n$, and $H_n,$ respectively,
still apply. The corresponding
zero-curvature equation reads
\begin{equation}\label{2.41}
   U_{t_n}-V_{n,x}+[U,V_n]=0, \quad n\in \mathbb{N}_0,
   \end{equation}
which results in the following set of equations
    \begin{align}
      & F_{n,x}=2G_n, \label{2.42c1} \\
      & G_{n,x}=-H_n +\left(-z^2\rho^2+z(u-u_{xx})+\frac{1}{4}\right)F_{n}, \label{2.42c2}\\
      & -2z\rho\rho_{t_n}+(u_{t_n}-u_{xxt_n})-H_{n,x}-
      2\left(-z^2\rho^2+z(u-u_{xx})+\frac{1}{4}\right)G_n=0.  \label{2.42c3}
    \end{align}
For fixed $n \in \mathbb{N}_0$, inserting the polynomial
expressions for $F_{n}$, $G_n$, and $H_n$ into (\ref{2.42c1})-(\ref{2.42c3}),
respectively, first yields recursion relations (\ref{2.3}) for
$f_{l}|_{l=0,\ldots,n}$, $g_{l}|_{l=0,\ldots,n}$,
$h_{l}|_{l=0,\ldots,n}$ and
      \begin{equation}\label{2.45}
        \begin{split}
        & h_{n+1}=-\frac{1}{2}f_{n-1,xx}+(u-u_{xx})f_{n}+\frac{1}{4}f_{n-1}, \\
        &   h_{n+2}=-\frac{1}{2}f_{n,xx}+\frac{1}{4}f_n.
        \end{split}
      \end{equation}
Moreover, using (\ref{2.42c3}), one finds
     \begin{equation}\label{2.46}
       \begin{split}
       &-2\rho\rho_{t_n}-h_{n+1,x}-(u-u_{xx})f_{n,x}-\frac{1}{4}f_{n-1,x}=0, \quad n\in \mathbb{N}_0,\\
    &   u_{t_n}-u_{xxt_n}-h_{n+2,x}-\frac{1}{4}f_{n,x}=0, \quad n\in \mathbb{N}_0.\\
       \end{split}
     \end{equation}
Hence, using (\ref{2.45}) and (\ref{2.46})
permits one to write the time-dependent  CHD2 hierarchy as
\begin{align}
        \textrm{CHD2}_n(u,\rho)&=
        \left(
          \begin{array}{c}
             -2\rho\rho_{t_n}-2(u-u_{xx})f_{n,x}-(u_x-u_{xxx})f_n-\frac{1}{2}f_{n-1,x}+\frac{1}{2}f_{n-1,xxx}\\
              u_{t_n}-u_{xxt_n}-\frac{1}{2}f_{n,x}+\frac{1}{2}f_{n,xxx} \\
          \end{array}
        \right)=0, \nonumber \\
      &  ~~~~~~~~~~~~~~~~~~~~~~~~~~~~~~~~~~~~~~~~~~~~~~~~~~~~~~
       n\in \mathbb{N}_0. \label{2.48}
      \end{align}
For convenience, we record the first equation in this hierarchy explicitly,
     \begin{equation}\label{2.49}
      \begin{split}
       &
       \mathrm{CHD2}_0(u,\rho)=
       \left(
         \begin{array}{c}
           -2\rho\rho_{t_0}+2u_{x}(u-u_{xx})+u(u_x-u_{xxx}) \\
           u_{t_0}-u_{xxt_0}+\frac{1}{2}u_x-\frac{1}{2}u_{xxx} \\
         \end{array}
       \right)=0.
       \end{split}
     \end{equation}
The first equation $\mathrm{CHD2}_0(u,\rho)=0$ in the
hierarchy is equivalent to the  CHD2 system as discussed in section
1, taking into account $u\rho=-2$ and the differential relation arising in the
$z^{n+2}$ term in (\ref{2.42c3}), that is, $-2u_x\rho^2=2u\rho\rho_x$.
Similarly, one can introduce the corresponding homogeneous CHD2
hierarchy by
 \begin{equation}\label{2.50}
        \widehat{\mathrm{CHD2}}_n(u,\rho)=\mathrm{CHD2}_n(u,\rho)|_{c_l=0,~l=1,\ldots,n}=0,
        \quad n\in\mathbb{N}_0.
    \end{equation}

\newtheorem{rem2.3}[hyp1]{Remark}
 \begin{rem2.3}
 The Lenard recursion formalism for the CHD2 hierarchy can be set up as follows.
 Define the two Lenard operators
 \begin{equation*}
   K=\left(
       \begin{array}{cc}
         \partial^3-\partial & 0 \\
         0 & -2\partial\rho^2-2\rho^2\partial \\
       \end{array}
     \right), \quad
     J=\left(
         \begin{array}{cc}
           2\partial m+2m\partial & -2\partial\rho^2-2\rho^2\partial \\
           -2\partial\rho^2-2\rho^2\partial & 0 \\
         \end{array}
       \right),
   \end{equation*}
 then the Lenard recursion sequence is given by
   \begin{equation*}
    K\vartheta_j=J\vartheta_{j+1}, \quad
    J\vartheta_0=0,
   \end{equation*}
 where $m=u-u_{xx}$, $ \vartheta_j=(f_j,f_{j+1})^T$, $j=0,\ldots,n-1$.
 Hence, using the zero-curvature equation $(\ref{2.41})$, one can obtain the
 CHD2 hierarchy.
 \end{rem2.3}

In fact, since the Lenard recursion formalism is almost universally
adopted in the contemporary literature, we thought it might be
worthwhile to use the Gesztesy's method, the polynomial recursion formalism,
to construct the CHD2 hierarchy.

\section{The stationary CHD2 formalism}
 This section is devoted to a detailed study of the stationary CHD2 hierarchy.
 We first
 define a fundamental meromorphic function $\phi(P,x)$ on the hyperelliptic
 curve $\mathcal{K}_n$, using the polynomial recursion formalism described in section 2,
 and then study the properties of the
 Baker-Akhiezer function $\psi(P,x,x_0)$, Dubrovin-type equations, and trace formulas.

For major parts of this section, we assume (\ref{2.1}), (\ref{2.3}),
(\ref{2.6}), (\ref{2.11})-(\ref{2.15c}), (\ref{2.27})-(\ref{2.30}),
and (\ref{2.34}), keeping $n\in\mathbb{N}_0$
fixed.

Recall the hyperelliptic curve $\mathcal{K}_n$
     \begin{equation}\label{3.1}
       \begin{split}
        & \mathcal{K}_n:  \mathcal{F}_n(z,y)=y^2-R_{2n+2}(z)=0, \\
        & R_{2n+2}(z)=-4 \prod_{m=0}^{2n+1} (z-E_m), \quad
      \{E_m\}_{m=0,\ldots,2n+1} \in \mathbb{C},
      \end{split}
     \end{equation}
which is compactified by joining two points at infinity
$P_{\infty_\pm}$, with $P_{\infty_+} \neq P_{\infty_-}$. But for
notational simplicity, the compactification is also denoted by
$\mathcal{K}_n$. Hence, $\mathcal{K}_n$ becomes a two-sheeted Riemann surface of
arithmetic genus $n$. Points $P$ on
      $\mathcal{K}_{n}\backslash\{P_{\infty\pm}\}$
are denoted by $P=(z,y(P))$, where $y(\cdot)$ is the
meromorphic function on $\mathcal{K}_{n}$ satisfying
       $\mathcal{F}_n(z,y(P))=0.$

The complex structure on $\mathcal{K}_{n}$ is defined in the usual
way by introducing local coordinates
$$\zeta_{Q_0}:P\rightarrow(z-z_0)$$
near points $Q_0=(z_0,y(Q_0))\in \mathcal{K}_{n},$
which are neither branch nor singular points of
$\mathcal{K}_{n}$; near the branch and singular points $Q_1=(z_1,y(Q_1)) \in \mathcal{K}_{n}$, the local coordinates are
   $$\zeta_{Q_1}:P \rightarrow (z-z_1)^{1/2};$$
near the points $P_{\infty_\pm} \in \mathcal{K}_{n}$, the local
coordinates are
   $$\zeta_{P_{\infty_\pm}}:P \rightarrow z^{-1}.$$

The holomorphic map
   $\ast,$ changing sheets, is defined by
       \begin{eqnarray}\label{3.2}
       && \ast: \begin{cases}
                        \mathcal{K}_{n}\rightarrow\mathcal{K}_{n},
                       \\
                       P=(z,y_j(z))\rightarrow
                       P^\ast=(z,y_{j+1(\mathrm{mod}~
                       2)}(z)), \quad j=0,1,
                      \end {cases}
                     \nonumber \\
      && P^{\ast \ast}:=(P^\ast)^\ast, \quad \mathrm{etc}.,
       \end{eqnarray}
where $y_j(z),\, j=0,1$ denote the two branches of $y(P)$ satisfying
$\mathcal{F}_{n}(z,y)=0$,  namely,
        \begin{equation}\label{3.17}
          (y-y_0(z))(y-y_1(z))
          =y^2-R_{2n+2}(z)=0.
        \end{equation}
Taking into account (\ref{3.17}), one easily finds
        \begin{equation}\label{3.18}
           \begin{split}
             & y_0+y_1=0,\\
             & y_0y_1=-R_{2n+2}(z),\\
             & y_0^2+y_1^2=2R_{2n+2}(z).\\
           \end{split}
        \end{equation}\\
Moreover, positive divisors on
$\mathcal{K}_{n}$ of degree $n$ are denoted by
        \begin{equation}\label{3.3}
          \mathcal{D}_{P_1,\ldots,P_{n}}:
             \begin{cases}
              \mathcal{K}_{n}\rightarrow \mathbb{N}_0,\\
              P\rightarrow \mathcal{D}_{P_1,\ldots,P_{n}}=
                \begin{cases}
                  \textrm{ $k$ \quad if $P$ occurs $k$
                      times in $\{P_1,\ldots,P_{n}\},$}\\
                   \textrm{ $0$ \quad if $P \notin
                     $$ \{P_1,\ldots,P_{n}\}.$}
                \end{cases}
             \end{cases}
        \end{equation}

Next, we define the stationary Baker-Akhiezer function
$\psi(P,x,x_0)$ on $\mathcal{K}_{n}\setminus \{P_{\infty_+},
P_{\infty_-},P_0=(0, y(0))\}$ by
       \begin{equation}\label{3.4}
         \begin{split}
          & \psi(P,x,x_0)=\left(
                            \begin{array}{c}
                              \psi_1(P,x,x_0) \\
                              \psi_2(P,x,x_0) \\
                            \end{array}
                          \right), \\
          & \psi_x(P,x,x_0)=U(u(x),\rho(x),z(P))\psi(P,x,x_0),\\
           & z^{-1}V_n(u(x),\rho(x),z(P))\psi(P,x,x_0)=y(P)\psi(P,x,x_0),\\
          & \psi_1(P,x_0,x_0)=1; \\
           &   P=(z,y)\in \mathcal{K}_{n}
           \setminus \{P_{\infty_+},P_{\infty_-},P_0=(0,y(0))\},~(x,x_0)\in \mathbb{R}^2.
         \end{split}
       \end{equation}
Closely related to $\psi(P,x,x_0)$ is the following meromorphic
function $\phi(P,x)$ on $\mathcal{K}_{n}$ defined by
       \begin{equation}\label{3.5}
         \phi(P,x)=
         \frac{ \psi_{1,x}(P,x,x_0)}{\psi_1(P,x,x_0)},
         \quad P\in \mathcal{K}_{n},~ x\in \mathbb{R}
       \end{equation}
such that
    \begin{equation}\label{3.6}
    \psi_1(P,x,x_0)=\mathrm{exp}\left(\int_{x_0}^x
         \phi(P,x^\prime)~ dx^\prime
         \right),
         \quad P\in \mathcal{K}_{n}\setminus \{P_{\infty_+}, P_{\infty_-},
         P_0\}.
    \end{equation}
Then, based on (\ref{3.4}) and (\ref{3.5}), a direct calculation
shows that
    \begin{align}\label{3.7}
        \phi(P,x)&=\frac{y+G_n(z,x)}{F_{n}(z,x)}
           \nonumber \\
        &=
         \frac{ H_n(z,x)}{y-G_n(z,x)},
    \end{align}
and
      \begin{equation}\label{3.8}
        \psi_2(P,x,x_0)= \psi_1(P,x,x_0)\phi(P,x).
      \end{equation}

In the following, the roots of polynomials $F_{n}$ and $H_n$ will
play a special role, and hence, we introduce on $\mathbb{C}\times\mathbb{R}$
 \begin{equation}\label{3.9}
         F_{n}(z,x)=f_0\prod_{j=1}^{n}(z-\mu_j(x)),
            \quad
            H_n(z,x)=h_0\prod_{l=1}^{n+2}(z-\nu_l(x)).
         \end{equation}
Moreover, we introduce
      \begin{equation}\label{3.10}
        \hat{\mu}_j(x)
            =(\mu_j(x),G_n(\mu_j(x),x))
            \in \mathcal{K}_{n}, \quad
            j=1,\ldots,n,~x\in\mathbb{R},
      \end{equation}
and
       \begin{equation}\label{3.11}
        \hat{\nu}_l(x)
            =(\nu_l(x),-G_n(\nu_l(x),x))
            \in \mathcal{K}_{n}, \quad
            l=1,\ldots,n+2,~x\in\mathbb{R}.
      \end{equation}
Due to assumption (\ref{2.1}), $u$ and $\rho$ are smooth and bounded, and hence,
$F_{n}(z,x)$ and $H_n(z,x)$ share the same property. Thus, one
concludes
    \begin{equation}\label{3.12}
        \mu_j,\nu_l \in C(\mathbb{R}), \quad
        j=1,\ldots,n,~l=1,\dots,n+2,
    \end{equation}
taking multiplicities (and appropriate reordering)
of the zeros of $F_{n}$ and $H_n$ into account.
From (\ref{3.7}),
the divisor $(\phi(P,x))$ of $\phi(P,x)$ is
given by
         \begin{equation}\label{3.14}
           (\phi(P,x))=\mathcal{D}_{\hat{\nu}_1(x)  \hat{\nu}_2(x) \underline{\hat{\nu}}(x)}(P)
           -\mathcal{D}_{P_{\infty_+} P_{\infty_-} \underline{\hat{\mu}}(x)}(P).
         \end{equation}
Here, we abbreviated
     \begin{equation}\label{3.15}
     \underline{\hat{\mu}}=\{\hat{\mu}_1,\ldots,\hat{\mu}_n\},
      \quad
      \underline{\hat{\nu}}=\{\hat{\nu}_3,\ldots,\hat{\nu}_{n+2}\}
      \in \mathrm{Sym}^n(\mathcal{K}_n).
     \end{equation}

Further properties of $\phi(P,x)$ are summarized as follows.
\newtheorem{lem3.1}{Lemma}[section]
 \begin{lem3.1}\label{lemma3.1}
    Suppose $(\ref{2.1})$, assume the $n$th stationary $CHD2$ equation $(\ref{2.34})$ holds, and
    let
    $P=(z,y)\in \mathcal{K}_{n}\setminus \{P_{\infty_+}, P_{\infty_-}\},$
    $(x,x_0) \in \mathbb{R}^2$. Then $\phi$ satisfies the Riccati-type equation
      \begin{equation}\label{3.19}
       \phi_x(P)+\phi(P)^2=-z^{2}\rho^2+z(u-u_{xx})+\frac{1}{4},
      \end{equation}
      as well as
       \begin{align}
        & \phi(P)\phi(P^\ast)=-\frac{H_n(z)}{F_{n}(z)},\label{3.20}
          \\
        &
         \phi(P)+\phi(P^\ast)=\frac{2G_n(z)}{F_{n}(z)}, \label{3.21}
          \\
        &
        \phi(P)-\phi(P^\ast)=\frac{2y}{F_{n}(z)}.\label{3.22}
       \end{align}
 \end{lem3.1}
\noindent
{\it Proof.}~Equation (\ref{3.19}) follows using the definition
 (\ref{3.7}) of $\phi$ as well as relations (\ref{2.15a})-(\ref{2.15c}).
Relations (\ref{3.20})-(\ref{3.22}) are clear from (\ref{2.19}), (\ref{3.18}), and (\ref{3.7}).
\quad $\square$

\vspace{0.1cm}
 The properties of $\psi(P,x,x_0)$ are summarized in the following lemma.
\newtheorem{lem3.2}[lem3.1]{Lemma}
 \begin{lem3.2}\label{lemma3.2}
    Suppose $(\ref{2.1})$, assume the $n$th stationary $CHD2$ equation $(\ref{2.34})$ holds, and
    let
    $P=(z,y)\in \mathcal{K}_{n}\setminus \{P_{\infty_+}, P_{\infty_-},P_0\},$
    $(x,x_0) \in \mathbb{R}^2$. Then $\psi_1(P,x,x_0), \psi_2(P,x,x_0)$ satisfy
      \begin{align}
       & \psi_1(P,x,x_0)=\Big(\frac{F_{n}(z,x)}{F_{n}(z,x_0)}\Big)^{1/2}
         \mathrm{exp}\left(y
          \int_{x_0}^x F_{n}(z,x^\prime)^{-1} dx^\prime
          \right), \label{3.26} \\
       &
          \psi_1(P,x,x_0)\psi_1(P^\ast,x,x_0)=\frac{F_{n}(z,x)}{F_{n}(z,x_0)}, \label{3.27}
          \\
       &
           \psi_2(P,x,x_0)\psi_2(P^\ast,x,x_0)=-\frac{H_n(z,x)}{ F_{n}(z,x_0)}, \label{3.28}
           \\
       &
           \psi_1(P,x,x_0)\psi_2(P^\ast,x,x_0)
        +\psi_1(P^\ast,x,x_0)\psi_2(P,x,x_0)
        =2\frac{G_n(z,x)}{F_{n}(z,x_0)}, \label{3.29}
        \\
       &
         \psi_1(P,x,x_0)\psi_2(P^\ast,x,x_0)
        -\psi_1(P^\ast,x,x_0)\psi_2(P,x,x_0)
        =\frac{-2y}{ F_{n}(z,x_0)}. \label{3.30}
      \end{align}
 \end{lem3.2}
 \noindent
{\it Proof.}~Equation (\ref{3.26}) is a consequence of (\ref{2.15a}), (\ref{3.6}), and (\ref{3.7}).
Equation (\ref{3.27}) is clear from (\ref{3.26}) and (\ref{3.28}) is a consequence of (\ref{3.8}),
(\ref{3.20}), and (\ref{3.27}). Equation (\ref{3.29}) follows using (\ref{3.8}), (\ref{3.21}), and
(\ref{3.27}). Finally, (\ref{3.30}) follows from (\ref{3.8}), (\ref{3.22}), and (\ref{3.27}).
\quad $\square$
\vspace{0.1cm}

In Lemma \ref{lemma3.2}, we denote by
      $$\psi_1(P)=\psi_{1,+}, ~ \psi_1(P^\ast)=\psi_{1,-},~
       \psi_2(P)=\psi_{2,+}, ~ \psi_2(P^\ast)=\psi_{2,-},$$
and then (\ref{3.27})-(\ref{3.30}) imply
     \begin{equation}\label{3.36}
        (\psi_{1,+}\psi_{2,-}-\psi_{1,-}\psi_{2,+})^2=
          (\psi_{1,+}\psi_{2,-}+\psi_{1,-}\psi_{2,+})^2
          -4\psi_{1,+}\psi_{2,-}\psi_{1,-}\psi_{2,+},
     \end{equation}
which is equivalent to the basic identity (\ref{2.19}),
$G_n^2+F_{n}H_n=R_{2n+2}$. This fact reveals the relations
between our approach and the algebro-geometric solutions of the CHD2
hierarchy.

\newtheorem{rem3.3}[lem3.1]{Remark}
  \begin{rem3.3}
  The Baker-Akhiezer function $\psi$ of the stationary $CHD2$ hierarchy
    is formally analogous to that defined
      in the context of KdV or AKNS
    hierarchies. However, its actual properties in a neighborhood
     of its essential singularity will feature characteristic differences
     to standard Baker-Akhiezer functions
     (cf. Remark $\ref{remark4.2}$).
  \end{rem3.3}

Next, we derive Dubrovin-type equations, that is,
first-order coupled systems of differential equations that govern the
dynamics of $\mu_j(x)$ and $\nu_l(x)$
with respect to variations of $x$.

\newtheorem{lem3.4}[lem3.1]{Lemma}
    \begin{lem3.4}\label{lemma3.4}
      Assume $(\ref{2.1})$ and the $n$th stationary $CHD2$ equation
   $(\ref{2.34})$ holds subject to the constraint $(\ref{2.26a})$.
   \begin{itemize}
      \item[\emph{(i)}]
     Suppose that the zeros $\{\mu_j(x)\}_{j=1,\ldots,n}$
     of $F_{n}(z,x)$ remain distinct for $ x \in
     \Omega_\mu,$ where $\Omega_\mu \subseteq \mathbb{R}$ is an open
     interval, then
     $\{\mu_j(x)\}_{j=1,\ldots,n}$ satisfy the system of
     differential equations,
    \begin{equation}\label{3.38}
           \mu_{j,x}= -2 \frac{ y(\hat{\mu}_j)}{f_0}
            \prod_{\scriptstyle k=1 \atop \scriptstyle k \neq j }^{n}
            (\mu_j(x)-\mu_k(x))^{-1}, \quad j=1,\ldots,n,
        \end{equation}
   with initial conditions
       \begin{equation}\label{3.39}
         \{\hat{\mu}_j(x_0)\}_{j=1,\ldots,n}
         \in \mathcal{K}_{n},
       \end{equation}
   for some fixed $x_0 \in \Omega_\mu$. The initial value
   problem $(\ref{3.38})$, $(\ref{3.39})$ has a unique solution
   satisfying
        \begin{equation}\label{3.40}
         \hat{\mu}_j \in C^\infty(\Omega_\mu,\mathcal{K}_{n}),
         \quad j=1,\ldots,n.
        \end{equation}
    \item[\emph{(ii)}]
    Suppose that the zeros $\{\nu_l(x)\}_{l=1,\ldots,n+2}$
     of $H_n(z,x)$ remain distinct for $ x \in
     \Omega_\nu,$ where $\Omega_\nu \subseteq \mathbb{R}$ is an open
     interval, then
     $\{\nu_l(x)\}_{l=1,\ldots,n+2}$ satisfy the system of
     differential equations,
    \begin{equation}\label{3.41}
           \nu_{l,x}=2\frac{(\nu_l^2\rho^2-(u-u_{xx})\nu_l-\frac{1}{4})y(\hat{\nu}_l)}{h_0}
            \prod_{\scriptstyle k=1 \atop \scriptstyle k \neq l }^{n+2}
            (\nu_l(x)-\nu_k(x))^{-1}, \quad l=1,\ldots,n+2,
        \end{equation}
   with initial conditions
       \begin{equation}\label{3.42}
         \{\hat{\nu}_l(x_0)\}_{l=1,\ldots,n+2}
         \in \mathcal{K}_{n},
       \end{equation}
   for some fixed $x_0 \in \Omega_\nu$. The initial value
   problem $(\ref{3.41})$, $(\ref{3.42})$ has a unique solution
   satisfying
        \begin{equation}\label{3.43}
         \hat{\nu}_l \in C^\infty(\Omega_\nu,\mathcal{K}_{n}),
         \quad l=1,\ldots,n+2.
        \end{equation}
   \end{itemize}
\end{lem3.4}
 \noindent
{\it Proof.}~It suffices to prove (\ref{3.38}) and (\ref{3.40})
 since the proof of (\ref{3.41}) and (\ref{3.43}) follow in
an identical manner. Differentiating (\ref{3.9}) with respect
to $x$ then yields
   \begin{equation}\label{3.44}
   F_{n,x}(\mu_j)= -f_0\mu_{j,x}
   \prod_{\scriptstyle k=1 \atop \scriptstyle k \neq j }^{n}
   (\mu_j(x)-\mu_k(x)).
   \end{equation}
On the other hand, taking into account equation (\ref{2.15a}), one finds
       \begin{equation}\label{3.45}
        F_{n,x}(\mu_j)=2 G_n(\mu_j)
        = 2 y(\hat{\mu}_j).
       \end{equation}
Then combining equation (\ref{3.44}) with (\ref{3.45}) leads to ({\ref{3.38}). The
proof of smoothness assertion (\ref{3.40}) is analogous to
the KdV case in \cite{15}. \quad $\square$
\vspace{0.1cm}

Next, we turn to the trace formulas of the CHD2 invariants, that
is, expressions of $f_{l}$ and $h_{l}$ in terms of symmetric
functions of the zeros $\mu_j$ and $\nu_l$ of $F_{n}$ and $H_n$,
respectively. For simplicity, we just record the simplest case.
\newtheorem{lem3.5}[lem3.1]{Lemma}
 \begin{lem3.5}\label{lemma3.5}
  Suppose $(\ref{2.1})$, assume the $n$th stationary $CHD2$ equation
     $(\ref{2.34})$ holds, and let $x\in\mathbb{R}.$ Then
    \begin{equation}\label{3.46}
    u^{-1}\mathcal{G}(-4u_x(u-u_{xx})-2u(u_x-u_{xxx}))
    = \sum_{j=1}^n \mu_j(x) -\frac{1}{2}\sum_{m=0}^{2n+1} E_m.
    \end{equation}
 \end{lem3.5}
\noindent
{\it Proof.}~Equation (\ref{3.46}) follows by considering the coefficient of $z^{n-1}$
in $F_{n}$ in (\ref{2.28}) and (\ref{3.9}),
 which yields
      \begin{equation}\label{3.47}
        \mathcal{G}(-4u_x(u-u_{xx})-2u(u_x-u_{xxx}))-c_1u=-f_0\sum_{j=1}^n \mu_j.
      \end{equation}
The constant $c_1$ can be determined by
considering the coefficient of $z^{2n+1}$ in (\ref{2.19}), which results in
     \begin{equation}\label{3.48}
        c_1=-\frac{1}{2}\sum_{m=0}^{2n+1} E_m.
     \end{equation}

\section{Stationary algebro-geometric solutions of CHD2 hierarchy}
 In this section, we obtain explicit Riemann theta function
 representations for the meromorphic function $\phi$,  the Baker-Akhiezer function $\psi$,
 and especially,
 for the solutions $u, \rho$ of the stationary CHD2 hierarchy.

We begin with the asymptotic properties of $\phi$ and
$\psi_j,j=1,2$.

\newtheorem{lem4.1}{Lemma}[section]
   \begin{lem4.1}\label{lemma4.1}
    Suppose $(\ref{2.1})$, assume the $n$th stationary $CHD2$ equation $(\ref{2.34})$
    holds, and let
    $P=(z,y)
    \in \mathcal{K}_n \setminus \{P_{\infty_+}, P_{\infty_-}, P_0\},$ $(x,x_0) \in
    \mathbb{R}^2$. Then
     \begin{align}
       & \phi(P) \underset{\zeta \rightarrow 0}{=}
         \pm i \rho \zeta^{-1}+\frac{\mp i (u-u_{xx})-\rho_x}{2\rho}+O(\zeta),
         \quad P \rightarrow P_{\infty_\pm}, \quad \zeta=z^{-1},
                \label{4.1} \\
       &
       \phi(P) \underset{\zeta \rightarrow 0}{=}
         \frac{1}{2}+(u-u_{x}) \zeta
          +O(\zeta^2),
           \quad P \rightarrow P_{0}, \quad \zeta=z,\label{4.2}
     \end{align}
and
     \begin{align}
      & \psi_1(P,x,x_0) \underset{\zeta \rightarrow 0}{=}
       \mathrm{exp} \left( \pm \frac{i}{\zeta} \int_{x_0}^x dx^\prime
          ~\rho(x^\prime)
          +O(1) \right),
        \quad
        P \rightarrow P_{\infty_\pm},\quad \zeta=z^{-1}, \label{4.3}
         \\
      &
      \psi_2(P,x,x_0) \underset{\zeta \rightarrow 0}{=}
        O(\zeta^{-1})~
        \mathrm{exp} \left( \pm \frac{i}{\zeta} \int_{x_0}^x dx^\prime
          ~\rho(x^\prime)
          +O(1) \right), \quad
P \rightarrow P_{\infty_\pm},\quad \zeta=z^{-1}, \label{4.4}
        \\
     &
      \psi_1(P,x,x_0) \underset{\zeta \rightarrow 0}{=}
      \mathrm{exp}\Big( \frac{1}{2}(x-x_0)\Big) (1+O(\zeta)),
                 \quad
           P \rightarrow P_{0}, \quad \zeta=z, \label{4.5}
          \\
     &
      \psi_2(P,x,x_0) \underset{\zeta \rightarrow 0}{=}
      \Big(\frac{1}{2}+O(\zeta)\Big)~
      \mathrm{exp}\Big( \frac{1}{2}(x-x_0)\Big) (1+O(\zeta)),
                 \quad
                    P \rightarrow P_{0}, \quad \zeta=z.  \label{4.6}
     \end{align}
  \end{lem4.1}
\noindent
  {\it Proof.}~The existence of the
asymptotic expansions of $\phi$ in terms of the appropriate
local coordinates $\zeta=z^{-1}$
near $P_{\infty_\pm}$ and $\zeta=z$ near $P_0$ is
clear from its explicit
expression in (\ref{3.7}). Next, we compute the coefficients
of these expansions utilizing the Riccati-type equation
 (\ref{3.19}). Indeed,
inserting the ansatz
\begin{equation}\label{4.7}
     \phi \underset{z \rightarrow \infty}{=}
     \phi_{-1}z +\phi_0 +O(z^{-1})
  \end{equation}
into (\ref{3.19}) and comparing the same powers of $z$ then yields
(\ref{4.1}). Similarly, inserting the ansatz
    \begin{equation}\label{4.8}
     \phi \underset{z \rightarrow 0}{=} \phi_0+
       \phi_{1} z +O(z^2)
    \end{equation}
into (\ref{3.19}) and comparing the same powers of $z$ then yields
(\ref{4.2}). Finally, expansions
(\ref{4.3})-(\ref{4.6}) follow from (\ref{3.6}), (\ref{3.8}), (\ref{4.1}), and
(\ref{4.2}). \quad $\square$

\newtheorem{rem4.2}[lem4.1]{Remark}
  \begin{rem4.2}\label{remark4.2}
      We note the fact
    that  $P_{\infty\pm}$, are the essential singularities
    of $\psi_j$, $j=1,2$.
    In addition, one easily finds the
    leading-order exponential term in $\psi_j$, $j=1,2,$ near $P_{\infty_\pm}$ is
    $x$-dependent, which makes matters worse.
    This is in sharp contrast to standard
    Baker-Akhiezer functions that typically feature a linear
    behavior
    with respect to $x$ in
    connection with their essential singularities of the type
    $\mathrm{exp}(c(x-x_0)\zeta^{-1})$
    near $\zeta=0$.
      \end{rem4.2}

Next, we introduce the holomorphic differentials $\eta_l(P)$ on
$\mathcal{K}_{n}$
      \begin{equation}\label{4.10}
        \eta_l(P)= \frac{z^{l-1}}{y(P)} dz,
        \quad l=1,\ldots,n,
      \end{equation}
and choose a homology basis $\{a_j,b_j\}_{j=1}^{n}$ on
$\mathcal{K}_{n}$ in such a way that the intersection matrix of the
cycles satisfies
$$a_j \circ b_k =\delta_{j,k},\quad a_j \circ a_k=0, \quad
b_j \circ   b_k=0, \quad j,k=1,\ldots, n.$$
Associated with $\mathcal{K}_n$, one
introduces an invertible
matrix $E \in GL(n, \mathbb{C})$
   \begin{equation}\label{4.11}
          \begin{split}
        & E=(E_{j,k})_{n \times n}, \quad E_{j,k}=
           \int_{a_k} \eta_j, \\
        &  \underline{c}(k)=(c_1(k),\ldots, c_{n}(k)), \quad
           c_j(k)=(E^{-1})_{j,k},
           \end{split}
   \end{equation}
and the normalized holomorphic differentials
        \begin{equation}\label{4.12}
          \omega_j= \sum_{l=1}^{n} c_j(l)\eta_l, \quad
          \int_{a_k} \omega_j = \delta_{j,k}, \quad
          \int_{b_k} \omega_j= \tau_{j,k}, \quad
          j,k=1, \ldots ,n.
        \end{equation}
Apparently, the matrix $\tau$ is symmetric and has a
positive-definite imaginary part.

We choose a fixed base point $Q_0 \in \mathcal{K}_{n} \setminus
\{P_{\infty_+}, P_{\infty_-}\}$.
The Abel
maps $\underline{A}_{Q_0}(\cdot) $ and
$\underline{\alpha}_{Q_0}(\cdot)$ are defined by
\begin{equation}\label{4.24}
    \begin{split}
    &
   \underline{A}_{Q_0}:\mathcal{K}_{n} \rightarrow
   J(\mathcal{K}_{n})=\mathbb{C}^{n}/L_{n}, \\
   &
   P \mapsto \underline{A}_{Q_0} (P)=(A_{Q_0,1}(P),\ldots,
  A_{Q_0,n} (P))
    =\left(\int_{Q_0}^P\omega_1,\ldots,\int_{Q_0}^P\omega_{n}\right)
  (\mathrm{mod}~L_{n})
  \end{split}
  \end{equation}
and
   \begin{equation}\label{4.25}
     \begin{split}
   & \underline{\alpha}_{Q_0}:
   \mathrm{Div}(\mathcal{K}_{n}) \rightarrow
   J(\mathcal{K}_{n}),\\
   & \mathcal{D} \mapsto \underline{\alpha}_{Q_0}
   (\mathcal{D})= \sum_{P\in \mathcal{K}_{n}}
    \mathcal{D}(P)\underline{A}_{Q_0} (P),
    \end{split}
    \end{equation}
where $L_{n}=\{\underline{z}\in \mathbb{C}^{n}|
           ~\underline{z}=\underline{N}+\tau\underline{M},
           ~\underline{N},~\underline{M}\in \mathbb{Z}^{n}\}.$

\vspace{0.15cm}

The following result shows the
nonlinearity of the Abel map with respect to
the variable $x$,
which
indicates a characteristic
difference between the CHD2 hierarchy and other completely integrable
systems such as the KdV and AKNS hierarchies.

\newtheorem{the4.3}[lem4.1]{Theorem}
  \begin{the4.3}\label{theorem4.3}
    Assume $(\ref{2.26a})$ and
    suppose that
    $\{\hat{\mu}_j(x)\}_{j=1,\ldots,n}$ satisfies the stationary
    Dubrovin equations $(\ref{3.38})$ on an open  interval $\Omega_\mu\subseteq\mathbb{R}$ such that
    $\mu_j(x),j=1,\ldots,n,$
    remain distinct and nonzero for $ x \in
      \Omega_\mu.$ Introducing the associated divisor
     $\mathcal{D}_{ \underline{\hat{\mu}}(x)} \in \mathrm{Sym}^n(\mathcal{K}_n)$, one computes
      \begin{equation}\label{4.15}
            \partial_x
            \underline{\alpha}_{Q_0}( \mathcal{D}_{ \underline{\hat{\mu}}(x)})
            = \frac{2}{ u(x)}
             \underline{c}(n),
            \qquad    x \in  \Omega_\mu.
        \end{equation}
   In particular, the Abel map does not linearize the divisor
   $\mathcal{D}_{\underline{\hat{\mu}}(x)}$ on $ \Omega_\mu$.
  \end{the4.3}
\noindent
{\it Proof.}~Let $x\in \Omega_\mu.$
 One finds
\begin{align}\label{4.18}
      \partial_x  \underline{\alpha}_{Q_0}(\mathcal{D}_{\underline{\hat{\mu}}(x)})
 &=
 \partial_x  \left(\sum_{j=1}^n \int_{Q_0}^{\hat{\mu}_j} \underline{\omega} \right)
 =\sum_{j=1}^n \mu_{j,x} \sum_{k=1}^n \underline{c}(k)
  \frac{\mu_j^{k-1}}{y(\hat{\mu}_j)}
      \nonumber \\
   &=
    \sum_{j=1}^n \sum_{k=1}^n
    -\frac{2}{f_0}
    \frac{\mu_j^{k-1}}
    {\prod_{\scriptstyle l=1 \atop \scriptstyle l \neq j }^{n}(\mu_j-\mu_l)}
    \underline{c}(k)
     \nonumber \\
    &=
      -\frac{2}{f_0}\sum_{k=1}^n \underline{c}(k) \delta_{k,n}
      =-\frac{2}{f_0}
        \underline{c}(n),
    \end{align}
where we used the  notation
 $\underline{\omega}=(\omega_1,\ldots,\omega_n)$,
 and a special case of Lagrange's interpolation formula (cf. Theorem E.1 \cite{15}),
   \begin{equation}\label{4.19}
   \sum_{j=1}^n \mu_j^{k-1} \prod_{\scriptstyle l=1 \atop \scriptstyle l \neq j}^n
   (\mu_j-\mu_l)^{-1}=\delta_{k,n}, \quad
   j,k=1,\ldots,n.
   \end{equation}

 The analogous results hold for the corresponding divisor
$\mathcal{D}_{\underline{\hat{\nu}}(x)}$ associated with
$\phi(P,x).$

\vspace{0.1cm}

Next, given the Riemann surface $\mathcal{K}_n$ and the homology basis $\{a_j,b_j\}_{j=1,\ldots,n}$,
one introduces the Riemann theta function by
$$\theta(\underline{z})=\sum_{\underline{n}\in\mathbb{Z}^n}\exp\Big(2\pi i(\underline{n},
\underline{z})+\pi i(\underline{n},\tau\underline{n})\Big),~~\underline{z}\in\mathbb{C}^n,$$
where $(\underline{A},\underline{B})=\sum_{j=1}^{n}\overline{A}_jB_j$ denotes the scalar product in $\mathbb{C}^n.$

Let
   \begin{equation}\label{4.26}
    \omega_{P_{\infty_+}, \hat{\nu}_1(x) }^{(3)}(P)=
    \frac{y-G_n(\nu_1)}{z-\nu_1}\frac{dz}{2y}
     -\frac{1}{2y} \prod_{j=1}^n(z-\lambda_j) dz
    \end{equation}
be the normalized differential of the third kind  holomorphic on
$\mathcal{K}_{n} \setminus \{P_{\infty_+}, \hat{\nu}_1(x)\}$ with simple poles at
$P_{\infty_+}$ and $\hat{\nu}_1(x)$  and residues $1$ and $-1$, respectively,
 \begin{align}
 & \omega_{P_{\infty_+},\hat{\nu}_1(x) }^{(3)}(P) \underset
              {\zeta \rightarrow 0}{=} (\zeta^{-1}+O(1))d \zeta,
              \quad \textrm{as $P \rightarrow P_{\infty_+},$}\label{4.27b}\\
             & \omega_{P_{\infty_+}, \hat{\nu}_1(x) }^{(3)}(P) \underset
              {\zeta \rightarrow 0}{=} (-\zeta^{-1}+O(1))d \zeta,
              \quad \textrm{as $P \rightarrow \hat{\nu}_1(x),$}\label{4.27a}
              \end{align}
where $\zeta$ in (\ref{4.27b}) denotes the local coordinate
    \begin{equation*}
        \zeta= z^{-1}
        ~~\textrm{for $P$ near $P_{\infty_+}$},
     \end{equation*}
     near $P_{\infty_+}$, and analogously, $\zeta$ in (\ref{4.27a}) that near $\hat{\nu}_1(x).$
The constants $\{\lambda_j\}_{j=1,\ldots,n}$ in (\ref{4.26}) are determined by
the normalization condition
      $$\int_{a_k} \omega_{P_{\infty_+}, \hat{\nu}_1(x) }^{(3)}=0,
      \qquad k=1,\ldots,n.$$

Similarly, let $\omega_{P_{\infty_-}, \hat{\nu}_2(x) }^{(3)}(P)$ be another
normalized differential of the third kind  holomorphic on
$\mathcal{K}_{n} \setminus \{P_{\infty_-}, \hat{\nu}_2(x)\}$ with simple poles at
$P_{\infty_-}$ and $\hat{\nu}_2(x)$  and residues $1$ and $-1$, respectively,
    \begin{align}
     & \omega_{P_{\infty_-}, \hat{\nu}_2(x) }^{(3)}(P)=
     \frac{y-G_n(\nu_2)}{z-\nu_2}\frac{dz}{2y}+
     \frac{1}{2y} \prod_{j=1}^n(z-\gamma_j) dz, \label{4.30a1} \\
     & \omega_{P_{\infty_-},\hat{\nu}_2(x) }^{(3)}(P) \underset
              {\zeta \rightarrow 0}{=} (\zeta^{-1}+O(1))d \zeta,
              \quad \textrm{as $P \rightarrow P_{\infty_-},$} \label{4.30a2} \\
             & \omega_{P_{\infty_-}, \hat{\nu}_2(x) }^{(3)}(P) \underset
              {\zeta \rightarrow 0}{=} (-\zeta^{-1}+O(1))d \zeta,
              \quad \textrm{as $P \rightarrow \hat{\nu}_2(x),$} \label{4.30a3}
    \end{align}
where $\zeta$ in (\ref{4.30a2}) denotes the local coordinate
    \begin{equation*}
        \zeta= z^{-1}
        ~~\textrm{for $P$ near $P_{\infty_-}$},
     \end{equation*}
near $P_{\infty_-}$, and analogously, $\zeta$ in (\ref{4.30a3}) that near $\hat{\nu}_2(x).$
The constants $\{\gamma_j\}_{j=1,\ldots,n}$ in (\ref{4.30a1}) are determined by
the normalization condition
      $$\int_{a_k} \omega_{P_{\infty_-}, \hat{\nu}_2(x) }^{(3)}=0,
      \qquad k=1,\ldots,n.$$
We define
    \begin{equation}\label{4.30a4}
    \Omega^{(3)}=\omega_{P_{\infty_+}, \hat{\nu}_1(x) }^{(3)}
    +\omega_{P_{\infty_-}, \hat{\nu}_2(x) }^{(3)}.
    \end{equation}
Then
   \begin{equation}\label{4.30a5}
     \int_{Q_0}^P \Omega^{(3)} \underset
  {\zeta \rightarrow 0}{=}  \mathrm{ln} \zeta +
  d_0+O(\zeta), \quad \textrm{as $P \rightarrow P_{\infty_\pm},$ }
   \end{equation}
for some constant $d_0 \in \mathbb{C}$.

Next, let $\omega_{P_{\infty_\pm},0}^{(2)}$ be normalized differentials of the second kind, satisfying
     \begin{align}
     & \int_{a_k} \omega_{P_{\infty_\pm},0}^{(2)}=0, \quad k=1,\ldots,n, \label{4.30a6}\\
     & \omega_{P_{\infty_\pm},0}^{(2)} \underset  {\zeta \rightarrow 0}{=}
        (\zeta^{-2} +O(1)) d\zeta, \quad \textrm{as $P \rightarrow P_{\infty_\pm}.$ } \label{4.30a7}
     \end{align}
We introduce
    \begin{equation}\label{4.30a8}
    \Omega_0^{(2)}=\omega_{P_{\infty_-},0}^{(2)}-\omega_{P_{\infty_+},0}^{(2)}.
    \end{equation}
Then
   \begin{equation}\label{4.30a9}
   \int_{Q_0}^P \Omega_0^{(2)} \underset  {\zeta \rightarrow 0}{=}
   \pm(\zeta^{-1}+e_{0,0}+O(\zeta)),
   \quad \textrm{as $P \rightarrow P_{\infty_\pm},$ }
   \end{equation}
for some constant $e_{0,0} \in \mathbb{C}$.
In the following, it will be convenient to introduce
the abbreviations
    \begin{eqnarray}\label{4.32}
         &&
           \underline{z}(P,\underline{Q})= \underline{\Xi}_{Q_0}
           -\underline{A}_{Q_0}(P)+\underline{\alpha}_{Q_0}
             (\mathcal{D}_{\underline{Q}}), \nonumber \\
          &&
           P\in \mathcal{K}_{n},\,
          \underline{Q}=(Q_1,\ldots,Q_{n})\in
          \mathrm{Sym}^{n}(\mathcal{K}_{n}),
         \end{eqnarray}
where $\underline{\Xi}_{Q_0}$ is
the vector of Riemann constants (cf.(A.45) \cite{15}).
It turns out that $\underline{z}(\cdot,\underline{Q}) $ is independent of the
choice of base point $Q_0$\,(cf.(A.52),\,(A.53) \cite{15}).

Based on above preparations, we will give
explicit representations for the meromorphic function $\phi$,
the Baker-Akhiezer function $\psi$,
and the
stationary CHD2 solutions $u, \rho$ in terms of the Riemann theta function
associated with $\mathcal{K}_{n}$.

\newtheorem{the4.5}[lem4.1]{Theorem}
 \begin{the4.5}\label{theorem4.5}
 Suppose
 $(\ref{2.1})$, and assume the
 $n$th stationary $CHD2$ equation
  $(\ref{2.34})$ holds on $ \Omega$ subject to the
  constraint $(\ref{2.26a})$. Moreover, let
$P=(z,y) \in \mathcal{K}_n \setminus \{P_{\infty_\pm}\}$ and $x,x_0 \in  \Omega$,
where $ \Omega \subseteq \mathbb{R}$ is an open interval. In
addition, suppose that $\mathcal{D}_{\underline{\hat{\mu}}(x)}$, or
equivalently, $\mathcal{D}_{\underline{\hat{\nu}}(x)}$ is nonspecial
for $x\in \Omega$. Then, $\phi$, $\psi$, $u,$ and $\rho$ admit the following
representations
  \begin{align}
  &  \phi(P,x)= i \rho(x)
    \frac{\theta(\underline{z}(P,\underline{\hat{\nu}}(x)))
            \theta(\underline{z}(P_{\infty_+},\underline{\hat{\mu}}(x)))}
            {\theta(\underline{z}(P_{\infty_+},\underline{\hat{\nu}}(x)))
            \theta(\underline{z}(P,\underline{\hat{\mu}}(x)))}
                   \mathrm{exp}\left(d_0
            -\int_{Q_0}^P \Omega^{(3)}\right),   \label{4.34}
       \\
  &
   \psi_1(P,x,x_0)=\frac{\theta(\underline{z}(P,\underline{\hat{\mu}}(x)))
            \theta(\underline{z}(P_{\infty_+},\underline{\hat{\mu}}(x_0)))}
            {\theta(\underline{z}(P_{\infty_+},\underline{\hat{\mu}}(x)))
            \theta(\underline{z}(P,\underline{\hat{\mu}}(x_0)))}
             \mathrm{exp}\left(
             \int_{x_0}^x dx^\prime~ i \rho(x^\prime) \int_{Q_0}^P \Omega_0^{(2)}
             \right), \label{4.34c1}
            \\
   &
   \psi_2(P,x,x_0)=i \rho(x)
            \frac{\theta(\underline{z}(P,\underline{\hat{\nu}}(x)))
            \theta(\underline{z}(P_{\infty_+},\underline{\hat{\mu}}(x_0)))}
            {\theta(\underline{z}(P_{\infty_+},\underline{\hat{\nu}}(x)))
            \theta(\underline{z}(P,\underline{\hat{\mu}}(x_0)))}
            \mathrm{exp}\left(d_0
            -\int_{Q_0}^P \Omega^{(3)}\right) \nonumber \\
      & ~~~~~~~~~~~~~~
       \times ~
            \mathrm{exp}\left(
             \int_{x_0}^x dx^\prime ~i \rho(x^\prime) \int_{Q_0}^P \Omega_0^{(2)}
             \right), \label{4.34c2}
            \\
       &
        u(x)=-4i \frac{\theta(\underline{z}(P_0,\underline{\hat{\nu}}(x)))
            \theta(\underline{z}(P_{\infty_+},\underline{\hat{\mu}}(x)))}
            {\theta(\underline{z}(P_{\infty_+},\underline{\hat{\nu}}(x)))
            \theta(\underline{z}(P_0,\underline{\hat{\mu}}(x)))}, \label{4.35}
            \\
       &
   \rho(x)=-\frac{i}{2}\frac{\theta(\underline{z}(P_0,\underline{\hat{\mu}}(x)))
            \theta(\underline{z}(P_{\infty_+},\underline{\hat{\nu}}(x)))}
            {\theta(\underline{z}(P_{\infty_+},\underline{\hat{\mu}}(x)))
            \theta(\underline{z}(P_0,\underline{\hat{\nu}}(x)))}.\label{4.35a1}
  \end{align}
\end{the4.5}
\noindent
{\it Proof.}~First, we temporarily assume that
    \begin{equation}\label{4.38}
      \mu_j(x)\neq \mu_{j^\prime}(x), \quad \nu_k(x)\neq \nu_{k^\prime}(x)
      \quad \textrm{for $j\neq j^\prime, k\neq k^\prime$ and
      $x\in\widetilde{\Omega}$},
      \end{equation}
 for appropriate $\widetilde{\Omega}\subseteq\Omega$. Since by (\ref{3.14}), $\mathcal
 {D}_{\hat{\nu}_1 \hat{\nu}_2  \underline{\hat{\nu}}}\sim
 \mathcal {D}_{P_{\infty_+} P_{\infty_-}  \underline{\hat{\mu}}}$, and
 $(P_{\infty_+}) ^\ast =P_{\infty_-} \notin\{\hat{\mu}_1,\ldots,\hat{\mu}_n \}$
 by hypothesis, one can use Theorem A.31 \cite{15} to
 conclude that $\mathcal {D}_{\underline{\hat{\nu}}}
 \in \textrm{Sym}^n(\mathcal {K}_n)$ is nonspecial. This
 argument is of course symmetric with respect to
 $\underline{\hat{\mu}}$ and $\underline{\hat{\nu}}$. Thus, $\mathcal
 {D}_{\underline{\hat{\mu}}}$ is nonspecial if and only
 if $\mathcal{D}_{\underline{\hat{\nu}}}$ is.

Next, we derive the representations of $\phi$, $\psi$, $u$, and $\rho$  in terms of the Riemann theta
function. A special case of Riemann's vanishing theorem (cf.\,Theorem A.26\,\cite{15})
yields
  \begin{equation}\label{4.39}
        \theta(\underline{\Xi}_{Q_0}-\underline{A}_{Q_0}(P)+\underline{\alpha}_{Q_0}(\mathcal
        {D}_{\underline{Q}}))=0 \quad \textrm{ if and only if $P\in
        \{Q_1,\ldots,Q_n\}$}.
  \end{equation}
Therefore, the divisor (\ref{3.14}) shows that $\phi(P,x)$ has expression of the
type
\begin{equation}\label{4.40}
C(x)\frac{
\theta(\underline{\Xi}_{Q_0}-\underline{A}_{Q_0}(P)+\underline{\alpha}_{Q_0}(\mathcal
{D}_{\underline{\hat{\nu}}(x)}))}{
\theta(\underline{\Xi}_{Q_0}-\underline{A}_{Q_0}(P)+\underline{\alpha}_{Q_0}(\mathcal
{D}_{\underline{\hat{\mu}}(x)}))} \mathrm{exp}\Big(d_0-\int_{Q_0}^P
\Omega^{(3)}
\Big),
\end{equation}
where $C(x)$ is independent of $P\in\mathcal {K}_n$. Then
 taking  into account
 the asymptotic expansion of $\phi(P,x)$ near $P_{\infty_+}$ in
(\ref{4.1}), we obtain (\ref{4.34}).

Now, let $\Psi$ denote the right-hand side of (\ref{4.34c1}). We intend to prove $\psi_1=\Psi$ with
$\psi_1$ given by (\ref{3.6}). For that purpose we first investigate the local
zeros and poles of $\psi_1$. Since they can
only come from zeros of $F_n(z,x^\prime)$ in (\ref{3.6}),
one computes using (\ref{3.10}), the definition (\ref{3.7}) of $\phi$,
and the Dubrovin equations (\ref{3.38}),
   \begin{align}
   \phi(P,x)&=\frac{y+G_n}{F_n}=\frac{y}{F_n}+\frac{F_{n,x}}{2F_n}
   \nonumber \\
    &=-\frac{1}{2}\frac{\mu_{j,x}}{z-\mu_j}-\frac{1}{2}\frac{\mu_{j,x}}{z-\mu_j}+O(1)
    \nonumber \\
    &=-\frac{\mu_{j,x}}{z-\mu_j}+O(1),
    \quad \textrm{as $ z \rightarrow \mu_j(x)$,} \label{4.40n1}
      \end{align}
where
  \begin{equation*}
  y \rightarrow y(\hat{\mu}_j(x))=G_n(\mu_j(x)),  \quad \textrm{as $ z \rightarrow \mu_j(x)$.}
  \end{equation*}
More concisely,
   \begin{equation}\label{4.40n2}
   \phi(P,x)=\frac{\partial }{\partial x} \mathrm{ln} (z-\mu_j(x)) +O(1)
   \quad \textrm{for $P$ near $\hat{\mu}_j(x)$},
   \end{equation}
which together with (\ref{3.6}) yields
   \begin{align}
   \psi_1(P,x,x_0)&=\mathrm{exp} \left(\int_{x_0}^x dx^\prime
   \left(\frac{\partial }{\partial x^\prime} \mathrm{ln} (z-\mu_j(x^\prime)) +O(1)
   \right) \right) \nonumber \\
   &=\frac{z-\mu_j(x)}{z-\mu_j(x_0)}O(1)
     \nonumber \\
   &=
   \left\{
     \begin{array}{ll}
       (z-\mu_j(x))O(1) & \hbox{for $P$ near $\hat{\mu}_j(x) \neq \hat{\mu}_j(x_0)$,} \\
       O(1) & \hbox{for $P$ near $\hat{\mu}_j(x) =\hat{\mu}_j(x_0)$,} \\
       (z-\mu_j(x_0))^{-1}O(1) & \hbox{for $P$ near $\hat{\mu}_j(x_0) \neq \hat{\mu}_j(x)$,}
     \end{array}
   \right. \label{4.40n3}
   \end{align}
where $O(1) \neq 0$ in (\ref{4.40n3}).
Consequently, $\psi_1$ and $\Psi$ have identical zeros and poles on $\mathcal{K}_n \setminus
\{P_{\infty_+}, P_{\infty_-}\}$, which are all simple by hypothesis (\ref{4.38}).
It remains to identify the behavior of $\psi_1$ and $\Psi$ near $P_{\infty_\pm}$.
Taking into account (\ref{3.6}), (\ref{4.3}), (\ref{4.30a9}),  and the expression (\ref{4.34c1}) for $\Psi$,
one observes that $\psi_1$ and $\Psi$ have identical exponential behavior up to order $O(1)$
near $P_{\infty_\pm}$. Thus, $\psi_1$ and $\Psi$ share the same
singularities and zeros, and the Riemann-Roch-type uniqueness result (cf. Lemma C.2 \cite{15})
then completes the proof $\psi_1=\Psi.$ The representation (\ref{4.34c2}) for $\psi_2$ on $ \widetilde{\Omega}$
follows using (\ref{3.8}), (\ref{4.34}), and (\ref{4.34c1}).
The representation (\ref{4.35a1}) for $\rho$ on $ \widetilde{\Omega}$
is clear from (\ref{4.2}) and (\ref{4.34}).
The representation
(\ref{4.35}) for $u$ on $ \widetilde{\Omega}$ follows from (\ref{4.35a1}) and the relation
$u\rho=-2$.
Finally, the extension of all these results from $x\in \widetilde{\Omega}$ to $x \in \Omega$
then simply follows from the continuity of $\underline{\alpha}_{O_0}$
and the hypothesis of $\mathcal{D}_{\underline{\hat{\mu}}(x)}$ being nonspecial for $x\in \Omega$.
\quad $\square$

\newtheorem{rem4.8}[lem4.1]{Remark}
  \begin{rem4.8}\label{remark4.8}
    Since by $(\ref{3.14})$ $\mathcal
 {D}_{ \hat{\nu}_1 \hat{\nu}_2 \underline{\hat{\nu}}}$ and
 $\mathcal {D}_{ P_{\infty_+} P_{\infty_-}\underline{\hat{\mu}}}$
 are linearly equivalent, that is
    \begin{equation}\label{4.42}
    \underline{A}_{Q_0}(P_{\infty_+})+\underline{A}_{Q_0}(P_{\infty_-}) + \underline{\alpha}_{Q_0}
    (\mathcal{D}_{\underline{\hat{\mu}}(x)})
    =\underline{A}_{Q_0}(\hat{\nu}_1(x))+\underline{A}_{Q_0}(\hat{\nu}_2(x))+ \underline{\alpha}_{Q_0}
    (\mathcal{D}_{\underline{\hat{\nu}}(x)}).
    \end{equation}
Then one infers
    \begin{equation}\label{4.43}
        \underline{\alpha}_{Q_0}
    (\mathcal{D}_{\underline{\hat{\mu}}(x)})
    =\underline{\Delta}+ \underline{\alpha}_{Q_0}
    (\mathcal{D}_{\underline{\hat{\nu}}(x)}),
    \quad \underline{\Delta}=\underline{A}_{P_{\infty_+}}(\hat{\nu}_1(x))+\underline{A}_{P_{\infty_-}}(\hat{\nu}_2(x)).
    \end{equation}
Hence, one can eliminate $\mathcal{D}_{\underline{\hat{\mu}}(x)}$
in $(\ref{4.34})$, $(\ref{4.35})$, and $(\ref{4.35a1})$, in terms of
$\mathcal{D}_{\underline{\hat{\nu}}(x)}$ using
   \begin{equation}\label{4.44}
    \underline{z}(P,\underline{\hat{\mu}})=
    \underline{z}(P,\underline{\hat{\nu}})+\underline{\Delta},
    \qquad P\in \mathcal{K}_n.
   \end{equation}
  \end{rem4.8}

\section{The time-dependent CHD2 formalism}
 In this section, we
 extend the algebro-geometric analysis of Section 3 to the
 time-dependent CHD2 hierarchy.

 Throughout this section, we assume (\ref{2.2}) holds.

 The time-dependent algebro-geometric initial value problem of the
 CHD2 hierarchy is to solve the time-dependent $r$th CHD2 flow with
 a stationary solution of the $n$th equation as initial data in the
 hierarchy. More precisely, given $n\in\mathbb{N}_0$, based on the
 solution $u^{(0)}, \rho^{(0)}$ of the $n$th stationary CHD2 equation
 $\textrm{s-CHD2}_n(u^{(0)},\rho^{(0)})=0$ associated with $\mathcal{K}_n$ and a
 set of integration constants $\{c_l\}_{l=1,\ldots,n} \subset
 \mathbb{C}$, we want to construct  a solution $u, \rho$ of the $r$th CHD2 flow
 $\mathrm{CHD2}_r(u,\rho)=0$ such that $u(t_{0,r})=u^{(0)},$  $\rho(t_{0,r})=\rho^{(0)}$ for some
 $t_{0,r} \in \mathbb{R},~r\in\mathbb{N}_0$.

To emphasize that the integration
 constants in the definitions of the stationary and the time-dependent CHD2
 equations are independent of each other, we indicate this by adding a tilde
 on all the time-dependent quantities. Hence, we employ the notation
 $\widetilde{V}_r,$ $\widetilde{F}_{r},$ $\widetilde{G}_r,$
 $\widetilde{H}_r,$ $\tilde{f}_{s}$, $\tilde{g}_{s},$ $\tilde{h}_{s}$, $\tilde{c}_s$
 in order to distinguish them from $V_n,$ $F_{n},$ $G_n,$ $H_n,$ $f_{l},$ $g_{l},$ $h_{l}$,
 $c_l$  in the following.
 In addition, we mark
 the individual $r$th CHD2 flow by a separate time variable $t_r \in
 \mathbb{R}$.

Summing up, we are seeking a solution $u, \rho$ of the time-dependent algebro-geometric initial
 value problem
   \begin{align}
    &
       \mathrm{CHD2}_r(u,\rho)=
        \left(
          \begin{array}{c}
           -2\rho \rho_{t_r}
            -2(u-u_{xx})\tilde{f}_{r,x}-(u_x-u_{xxx})\tilde{f}_{r}
          -\frac{1}{2} \tilde{f}_{r-1,x}+\frac{1}{2}\tilde{f}_{r-1,xxx} \\
            u_{t_r}-u_{xxt_r}-\frac{1}{2} \tilde{f}_{r,x}+\frac{1}{2}\tilde{f}_{r,xxx} \\
          \end{array}
        \right)=0, \label{5.1}
         \\
       & (u,\rho)|_{t_r=t_{0,r}}=(u^{(0)},\rho^{(0)}), \nonumber \\
     &
      \textrm{s-CHD2}_n(u^{(0)},\rho^{(0)})=
     \left(
       \begin{array}{c}
         -2(u-u_{xx})f_{n,x}-(u_x-u_{xxx})f_{n}
          -\frac{1}{2} f_{n-1,x}+\frac{1}{2} f_{n-1,xxx}  \\
         -\frac{1}{2}f_{n,x}+\frac{1}{2} f_{n,xxx} \\
       \end{array}
     \right)=0, \label{5.1a8}
    \end{align}
for some $t_{0,r}\in\mathbb{R},$ $n,r\in\mathbb{N}_0$, where $u=u(x,t_r),$ $\rho=\rho(x,t_r)$
 satisfy (\ref{2.2}), and the curve $\mathcal{K}_n$ is
 associated with the initial data $(u^{(0)}, \rho^{(0)})$ in (\ref{5.1a8}). Noticing that
 the CHD2 flows are isospectral, we further
 assume that (\ref{5.1a8}) holds not only for $t_r=t_{0,r}$, but also for all $t_r \in
 \mathbb{R}$. Hence, we start with
 the zero-curvature equations
    \begin{equation}\label{5.3}
        U_{t_r}-\widetilde{V}_{r,x}+[U,\widetilde{V}_r]=0,
    \end{equation}
    \begin{equation}\label{5.4}
        -V_{n,x}+[U,V_n]=0,
    \end{equation}
where
 \begin{equation}\label{5.5}
    \begin{split}
    & U(z)=
     \left(
       \begin{array}{cc}
         0 & 1 \\
         -z^{2}\rho^2+z(u-u_{xx})+\frac{1}{4} & 0 \\
       \end{array}
     \right),
     \\
    & V_n(z)=
    \left(
      \begin{array}{cc}
      -zG_n(z) & zF_{n}(z) \\
      zH_n(z) & zG_n(z) \\
      \end{array}
    \right),
      \\
    & \widetilde{V}_r(z)=
       \left(
         \begin{array}{cc}
           -z\widetilde{G}_r(z) & z\widetilde{F}_{r}(z) \\
           z\widetilde{H}_r(z) & z\widetilde{G}_r(z) \\
         \end{array}
       \right),
    \end{split}
  \end{equation}
and
  \begin{eqnarray}\label{5.6}
   &&
    F_{n}(z)=\sum_{l=0}^{n} f_{l} z^{n-l}
    =f_0\prod_{j=1}^n (z-\mu_j),
    \\
   &&
    G_n(z)=\sum_{l=0}^{n} g_{l} z^{n-l},
     \\
   &&
    H_n(z)=\sum_{l=0}^{n+2} h_{l} z^{n+2-l}
    =h_0 \prod_{l=1}^{n+2} (z-\nu_l),\\
   &&
    \widetilde{F}_{r}(z)=\sum_{s=0}^{r} \tilde{f}_{s} z^{r-s},
    \\
   &&
    \widetilde{G}_r(z)=\sum_{s=0}^{r} \tilde{g}_{s} z^{r-s},
    \\
    &&
    \widetilde{H}_r(z)=\sum_{s=0}^{r+2} \tilde{h}_{s} z^{r+2-s},
   \end{eqnarray}
for fixed $n,r\in \mathbb{N}_0$. Here, $\{f_{l}\}_{l=0,\ldots,n},$
 $\{g_{l}\}_{l=0,\ldots,n}$, $\{h_{l}\}_{l=0,\ldots,n+2}$,
  $\{\tilde{f}_{s}\}_{s=0,\ldots,r},$
 $\{\tilde{g}_{s}\}_{s=0,\ldots,r}$, and $\{\tilde{h}_{s}\}_{s=0,\ldots,r+2}$
 are defined as in
 (\ref{2.3}), with $u(x), \rho(x)$ replaced by $u(x,t_r), \rho(x,t_r)$ etc., and with appropriate
 integration constants.
Explicitly, (\ref{5.3}) and (\ref{5.4}) are equivalent to
  \begin{eqnarray}
        &&
      -2z\rho \rho_{t_r}+(u_{t_r}-u_{xxt_r})-\widetilde{H}_{r,x}
        -2\Big(-z^2\rho^2+z(u-u_{xx})+\frac{1}{4}\Big)\widetilde{G}_r=0, \nonumber \\ \label{5.12}
        \\
      &&
       \widetilde{F}_{r,x} =2\widetilde{G}_r,\label{5.13} \\
      &&
      \widetilde{G}_{r,x}=-\widetilde{H}_r
        + \Big(-z^2\rho^2+z(u-u_{xx})+\frac{1}{4}\Big)\widetilde{F}_{r}\label{5.14}
    \end{eqnarray}
and
   \begin{align}
    &
     F_{n,x}=2G_n, \label{5.15} \\
    &
      H_{n,x}=-2\Big(-z^2\rho^2+z(u-u_{xx})+\frac{1}{4}\Big)G_n, \label{5.16} \\
    &
    G_{n,x}=-H_n +\Big(-z^2\rho^2+z(u-u_{xx})+\frac{1}{4}\Big)F_{n}.\label{5.17}
   \end{align}
From (\ref{5.15})-(\ref{5.17}), one finds
  \begin{equation}\label{5.18}
    \frac{d}{dx} \mathrm{det}(V_n(z))=-z^2\frac{d}{dx}
    \Big( G_n(z)^2+F_{n}(z)H_n(z) \Big)=0,
  \end{equation}
and meanwhile (see Lemma \ref{lemma5.2})
  \begin{equation}\label{5.19}
    \frac{d}{dt_r} \mathrm{det}(V_n(z))=-z^2\frac{d}{dt_r}
    \Big( G_n(z)^2+F_{n}(z)H_n(z) \Big)=0.
  \end{equation}
Hence, $G_n(z)^2+F_{n}(z)H_n(z)$ is independent of variables
both $x$ and $t_r$, which implies the fundamental identity (\ref{2.19}) holds,
   \begin{equation}\label{5.20}
   G_n(z)^2+F_{n}(z)H_n(z)=R_{2n+2}(z),
   \end{equation}
and the hyperelliptic curve
$\mathcal{K}_n$ is still given by (\ref{2.27}).

Next, we define the time-dependent Baker-Akhiezer function
$\psi(P,x,x_0,t_r,t_{0,r})$ on
$\mathcal{K}_n \setminus \{ P_{\infty_\pm},P_0 \}$ by
  \begin{equation}\label{5.21}
         \begin{split}
          & \psi(P,x,x_0,t_r,t_{0,r})=\left(
                            \begin{array}{c}
                              \psi_1(P,x,x_0,t_r,t_{0,r}) \\
                              \psi_2(P,x,x_0,t_r,t_{0,r}) \\
                            \end{array}
                          \right), \\
          & \psi_x(P,x,x_0,t_r,t_{0,r})=U(u(x,t_r), \rho(x,t_r), z(P))\psi(P,x,x_0,t_r,t_{0,r}),\\
          &  \psi_{t_r}(P,x,x_0,t_r,t_{0,r})=z^{-1}\widetilde{V}_r(u(x,t_r),\rho(x,t_r),z(P))
               \psi(P,x,x_0,t_r,t_{0,r}), \\
           & z^{-1}V_n(u(x,t_r),\rho(x,t_r),z(P))\psi(P,x,x_0,t_r,t_{0,r})
           =y(P)\psi(P,x,x_0,t_r,t_{0,r}),\\
          & \psi_1(P,x_0,x_0,t_{0,r},t_{0,r})=1; \\
          &
          P=(z,y)\in \mathcal{K}_{n}
           \setminus \{P_{\infty_\pm},P_0\},~(x,t_r)\in
           \mathbb{R}^2.
         \end{split}
   \end{equation}
Closely related to $\psi(P,x,x_0,t_r,t_{0,r})$ is the following
meromorphic function $\phi(P,x,t_r)$ on $\mathcal{K}_{n}$ defined by
        \begin{equation}\label{5.23}
         \phi(P,x,t_r)=
         \frac{ \psi_{1,x}(P,x,x_0,t_r,t_{0,r})}
          {\psi_1(P,x,x_0,t_r,t_{0,r})},
                      \quad P \in
          \mathcal{K}_{n}\setminus \{P_{\infty_\pm},P_0\},
          ~ (x,t_r)\in \mathbb{R}^2
        \end{equation}
such that
\begin{align}\label{5.22}
   \psi_1(P,x,x_0,t_r,t_{0,r})&=\mathrm{exp}\Big(
     \int_{t_{0,r}}^{t_r} ds
     (\widetilde{F}_{r}(z,x_0,s)\phi(P,x_0,s)
          -\widetilde{G}_r(z,x_0,s))
      \nonumber \\
     &+  \int_{x_0}^x dx^\prime
     \phi(P,x^\prime,t_r)
      \Big),
     \quad
     P=(z,y)\in \mathcal{K}_{n}
           \setminus \{P_{\infty_\pm},P_0\}.
   \end{align}
Then, using (\ref{5.21}) and (\ref{5.23}),  one infers that
   \begin{align}\label{5.24}
        \phi(P,x,t_r)&= \frac{y+G_n(z,x,t_r)}{F_{n}(z,x,t_r)}
           \nonumber \\
        &=
         \frac{H_n(z,x,t_r)}{y-G_n(z,x,t_r)},
    \end{align}
 and
      \begin{equation}\label{5.25}
        \psi_2(P,x,x_0,t_r,t_{0,r})=
        \psi_1(P,x,x_0,t_r,t_{0,r})\phi(P,x,t_r).
      \end{equation}
In analogy to (\ref{3.10}) and (\ref{3.11}), we introduce
     \begin{align}
      & \hat{\mu}_j(x,t_r)=(\mu_j(x,t_r),
       G_n(\mu_j(x,t_r),x,t_r))
       \in \mathcal{K}_n,   \quad
        j=1,\ldots,n, ~(x,t_r)\in \mathbb{R}^2,  \label{5.26}\\
      &
      \hat{\nu}_l(x,t_r)=(\nu_l(x,t_r),
       -G_n(\nu_l(x,t_r),x,t_r))
       \in \mathcal{K}_n,  \quad
       l=1,\ldots,n+2, ~(x,t_r)\in \mathbb{R}^2.  \label{5.27}
     \end{align}
The regularity properties of $F_{n}$, $H_n$, $\mu_j$, and $\nu_l$ are
analogous to those in Section 3 due to assumptions
(\ref{2.2}).
Similar to (\ref{3.14}), the divisor $(\phi(P,x,t_r))$ of
$\phi(P,x,t_r)$ reads
 \begin{equation}\label{5.28}
           (\phi(P,x,t_r))=
           \mathcal{D}_{\hat{\nu}_1(x,t_r) \hat{\nu}_2(x,t_r) \underline{\hat{\nu}}(x,t_r)}(P)
           -\mathcal{D}_{P_{\infty_+} P_{\infty_-} \underline{\hat{\mu}}(x,t_r)}(P)
  \end{equation}
with
  \begin{equation}\label{5.29}
    \underline{\hat{\mu}}=\{\hat{\mu}_1,\ldots,\hat{\mu}_{n}\},
    \quad
    \underline{\hat{\nu}}=\{\hat{\nu}_3,\ldots,\hat{\nu}_{n+2}\}
    \in \mathrm{Sym}^n (\mathcal{K}_n).
  \end{equation}

The properties of $\phi(P,x,t_r)$ are summarized as follows.
\newtheorem{lem5.1}{Lemma}[section]
 \begin{lem5.1}
  Assume $(\ref{2.2})$ and suppose that $(\ref{5.3})$,
    $(\ref{5.4})$ hold. Moreover, let
    $P=(z,y) \in \mathcal{K}_{n}\setminus \{P_{\infty_\pm},P_0\}$ and
    $(x,t_r)\in \mathbb{R}^2.$ Then $\phi$ satisfies
    \begin{align}
    &
    \phi_x(P)+ \phi(P)^2 =-z^{2}\rho^2 +z(u-u_{xx})+\frac{1}{4},\label{5.30} \\
    &
    \phi_{t_r}(P)= (-\widetilde{G}_r(z)+\widetilde{F}_{r}(z)\phi(P))_x \label{5.31} \\
    &~~~~~~~~
             = \widetilde{H}_r(z)+\Big(z^{2}\rho^2-z(u- u_{xx})-\frac{1}{4}\Big)\widetilde{F}_{r}(z)
                +(\widetilde{F}_{r}(z)\phi(P))_x, \nonumber \\
    &
    \phi_{t_r}(P)=\widetilde{H}_r(z)+2\widetilde{G}_r(z)\phi(P)
         -\widetilde{F}_{r}(z)\phi(P)^2,\label{5.32}\\
         &\phi(P)\phi(P^\ast)=-\frac{H_n(z)}{F_{n}(z)},\label{5.33}\\
         &\phi(P)+\phi(P^\ast)=\frac{2G_n(z)}{F_{n}(z)},\label{5.34}\\
         & \phi(P)-\phi(P^\ast)=\frac{2y}{F_{n}(z)}.\label{5.35}
   \end{align}
\end{lem5.1}
\noindent
{\it Proof.}~Equations (\ref{5.30}) and (\ref{5.33})-(\ref{5.35})
can be proved as in Lemma \ref{lemma3.1}. Using (\ref{5.21}) and
(\ref{5.23}), one infers that
\begin{align}\label{5.36}
       & \phi_{t_r}=
        (\mathrm{ln}\psi_1)_{xt_r}=(\mathrm{ln}\psi_1)_{t_rx}
        =\Big(\frac{\psi_{1,t_r}}{\psi_1}\Big)_x
          \nonumber \\
      &~~~~
  =
 \Big(\frac{-\widetilde{G}_r\psi_1+\widetilde{F}_{r}\psi_2}{\psi_1}\Big)_x
     = (-\widetilde{G}_r+\widetilde{F}_{r}\phi)_x.
 \end{align}
 Insertion of (\ref{5.14}) into
(\ref{5.36}) then yields (\ref{5.31}).
To prove (\ref{5.32}),
 one observes that
 \begin{align}\label{5.37}
     \phi_{t_r} =&  \Big(\frac{\psi_2}{\psi_1}\Big)_{t_r}
     =
       \Big(\frac{\psi_{2,t_r}}{\psi_1}-
      \frac{\psi_2\psi_{1,t_r}}{\psi_1^2} \Big)
        \nonumber \\
     =&
      \Big(\frac{\widetilde{H}_r\psi_1+\widetilde{G}_r\psi_2}{\psi_1}
     -\phi\frac{-\widetilde{G}_r\psi_1+\widetilde{F}_{r}\psi_2}{\psi_1}\Big)
       \nonumber \\
     =&
     \widetilde{H}_r+2\widetilde{G}_r\phi-\widetilde{F}_{r}\phi^2,
 \end{align}
which leads to (\ref{5.32}). Alternatively, one can also insert (\ref{5.12})-(\ref{5.14})
into (\ref{5.31}) to obtain (\ref{5.32}). \quad $\square$ \\

Next, we determine the time evolution of $F_{n}$, $G_n$, and $H_n$,
using relations (\ref{5.12})-(\ref{5.14}) and (\ref{5.15})-(\ref{5.17}).
\newtheorem{lem5.2}[lem5.1]{Lemma}
 \begin{lem5.2}\label{lemma5.2}
  Assume $(\ref{2.2})$ and suppose that $(\ref{5.3})$,
  $(\ref{5.4})$ hold. Then
   \begin{align}
    & F_{n,t_r}=2(G_n\widetilde{F}_{r}-\widetilde{G}_rF_{n}), \label{5.38} \\
    &  G_{n,t_r}=\widetilde{H}_rF_{n}-H_n\widetilde{F}_{r}, \label{5.39} \\
    &  H_{n,t_r}=2(H_n\widetilde{G}_r-G_n\widetilde{H}_r). \label{5.40}
   \end{align}
 Equations $(\ref{5.38})$--$(\ref{5.40})$ are equivalent to
    \begin{equation}\label{5.41}
        -V_{n,t_r}+[z^{-1}\widetilde{V}_r,V_n]=0.
    \end{equation}
 \end{lem5.2}
\noindent
{\it Proof.}~Differentiating (\ref{5.35}) with
respect to $t_r$ naturally yields
  \begin{equation}\label{5.42}
    (\phi(P)-\phi(P^\ast))_{t_r}=-2yF_{n,t_r}F_{n}^{-2}.
  \end{equation}
On the other hand, using (\ref{5.32}), (\ref{5.34}), and (\ref{5.35}),
the left-hand side of (\ref{5.42}) can be expressed as
   \begin{align}\label{5.43}
    \phi(P)_{t_r}-\phi(P^\ast)_{t_r}=&
    2\widetilde{G}_r(\phi(P)-\phi(P^\ast))-\widetilde{F}_{r}
    (\phi(P)^2-\phi(P^\ast)^2)
     \nonumber \\
    =&
    4y(\widetilde{G}_rF_{n}-\widetilde{F}_{r}G_n)F_{n}^{-2}.
   \end{align}
Combining (\ref{5.42}) and (\ref{5.43}) then proves (\ref{5.38}).
Similarly, differentiating (\ref{5.34}) with respect
to $t_r$, one finds
  \begin{equation}\label{5.44}
    (\phi(P)+\phi(P^\ast))_{t_r}=2(G_{n,t_r}F_{n}-G_nF_{n,t_r})F_{n}^{-2}.
  \end{equation}
Meanwhile, the
left-hand side of (\ref{5.44})  also equals
  \begin{align}\label{5.45}
  \phi(P)_{t_r}+\phi(P^\ast)_{t_r}=&
   2\widetilde{G}_r(\phi(P)+\phi(P^\ast))
   -\widetilde{F}_{r}(\phi(P)^2+\phi(P^\ast)^2)+2\widetilde{H}_r
    \nonumber \\
   =&
   -2G_nF_{n}^{-2}F_{n,t_r}
   +2F_{n}^{-1}(\widetilde{H}_rF_{n}-\widetilde{F}_{r}H_n),
  \end{align}
  using (\ref{5.32}), (\ref{5.33}), and (\ref{5.34}).
Equation (\ref{5.39}) is  clear from
(\ref{5.44}) and (\ref{5.45}). Then, (\ref{5.40}) follows by
differentiating (\ref{2.19}), that is, $G_n^2+F_{n}H_n=R_{2n+2}(z)$,
with respect to $t_r$, and using  (\ref{5.38}) and (\ref{5.39}).
Finally,
a direct calculation shows (\ref{5.41}) holds. \quad $\square$
\vspace{0.1cm}

Basic properties of $\psi(P,x,x_0,t_r,t_{0,r})$ are summarized as follows.

\newtheorem{lem5.3}[lem5.1]{Lemma}
 \begin{lem5.3}
  Assume $(\ref{2.2})$ and suppose that $(\ref{5.3})$,
  $(\ref{5.4})$ hold.  Moreover, let
  $P=(z,y) \in \mathcal{K}_{n}\setminus \{P_{\infty_\pm},P_0\}$ and
   $(x,x_0,t_r,t_{0,r})\in \mathbb{R}^4.$
  Then, the Baker-Akhiezer function $\psi$ satisfies
  \begin{align}
   & \psi_1(P,x,x_0,t_r,t_{0,r}) =
    \Big(\frac{F_{n}(z,x,t_r)}{F_{n}(z,x_0,t_{0,r})}\Big)^{1/2}
    \mathrm{exp} \Bigg(
    y \int_{t_{0,r}}^{t_r} ds
    \widetilde{F}_{r}(z,x_0,s)F_{n}(z,x_0,s)^{-1}
    \nonumber \\
    &~~~~~~~~~~~~~~~~~~~~~~~~~~
    +y \int_{x_0}^x dx^\prime
    F_{n}(z,x^\prime,t_r)^{-1}
    \Bigg),\label{5.46} \\
  &  \psi_1(P,x,x_0,t_r,t_{0,r}) \psi_1(P^\ast,x,x_0,t_r,t_{0,r})
    =\frac{F_{n}(z,x,t_r)}{F_{n}(z,x_0,t_{0,r})}, \label{5.47} \\
  & \psi_2(P,x,x_0,t_r,t_{0,r}) \psi_2(P^\ast,x,x_0,t_r,t_{0,r})
    =-\frac{H_n(z,x,t_r)}{ F_{n}(z,x_0,t_{0,r})}, \label{5.48} \\
  & \psi_1(P,x,x_0,t_r,t_{0,r}) \psi_2(P^\ast,x,x_0,t_r,t_{0,r})
    + \psi_1(P^\ast,x,x_0,t_r,t_{0,r}) \psi_2(P,x,x_0,t_r,t_{0,r})
    =2\frac{G_n(z,x,t_r)}{F_{n}(z,x_0,t_{0,r})},\label{5.49} \\
  & \psi_1(P,x,x_0,t_r,t_{0,r}) \psi_2(P^\ast,x,x_0,t_r,t_{0,r})
    - \psi_1(P^\ast,x,x_0,t_r,t_{0,r}) \psi_2(P,x,x_0,t_r,t_{0,r})
    =-\frac{2y}{ F_{n}(z,x_0,t_{0,r})}.\label{5.50}
  \end{align}
\end{lem5.3}
\noindent
{ \it Proof.}~To prove (\ref{5.46}), we first
consider the part of time variable in the definition (\ref{5.22}),
that is,
   \begin{equation}\label{5.51}
    \mathrm{exp} \left( \int_{t_{0,r}}^{t_r} ds~
    (\widetilde{F}_{r}(z,x_0,s)\phi(P,x_0,s)
    -\widetilde{G}_r(z,x_0,s))
    \right).
   \end{equation}
The integrand in the above integral equals
   \begin{align}\label{5.52}
   &
     \widetilde{F}_{r}(z,x_0,s)\phi(P,x_0,s)
     -\widetilde{G}_r(z,x_0,s)
     \nonumber \\
   &~~~~~
     =\widetilde{F}_{r}(z,x_0,s)
     \frac{y+G_n(z,x_0,s)}{F_{n}(z,x_0,s)}
     -\widetilde{G}_r(z,x_0,s)
       \nonumber \\
     &~~~~~
     =y \widetilde{F}_{r}(z,x_0,s)F_{n}(z,x_0,s)^{-1}
      +(\widetilde{F}_{r}(z,x_0,s)G_n(z,x_0,s)
       \nonumber \\
    &~~~~~~~~
     -\widetilde{G}_r(z,x_0,s)F_{n}(z,x_0,s))F_{n}(z,x_0,s)^{-1}
       \nonumber \\
      &~~~~~
      =y \widetilde{F}_{r}(z,x_0,s)F_{n}(z,x_0,s)^{-1}
      +\frac{1}{2}\frac{F_{n,s}(z,x_0,s)}{F_{n}(z,x_0,s)},
   \end{align}
using (\ref{5.24}) and (\ref{5.38}). Hence,
(\ref{5.51}) can be expressed as
   \begin{equation}\label{5.53}
   \Big(\frac{F_{n}(z,x_0,t_r)}{F_{n}(z,x_0,t_{0,r})}\Big)^{1/2}
     \mathrm{exp} \left( y \int_{t_{0,r}}^{t_r} ds
     \widetilde{F}_{r}(z,x_0,s)F_{n}(z,x_0,s)^{-1}
     \right).
   \end{equation}
On the other hand, the
part of space variable in (\ref{5.22}) can be written as
 \begin{equation}\label{5.54}
    \Big(\frac{F_{n}(z,x,t_r)}{F_{n}(z,x_0,t_{r})}\Big)^{1/2}
    \mathrm{exp}\left(y \int_{x_0}^x dx^\prime
    F_{n}(z,x^\prime,t_r)^{-1}  \right),
 \end{equation}
 using the similar procedure in Lemma \ref{lemma3.2}. Then
combining (\ref{5.53}) and (\ref{5.54}) readily leads to (\ref{5.46}).
Evaluating (\ref{5.46}) at the points $P$ and $P^\ast$
and multiplying the resulting expressions yields (\ref{5.47}).
The remaining statements are direct consequences of (\ref{5.25}),
(\ref{5.33})-(\ref{5.35}), and (\ref{5.47}). \quad $\square$

\vspace{0.1cm}

In analogy to Lemma \ref{lemma3.4}, the dynamics of the zeros
$\{\mu_j(x,t_r)\}_{j=1,\ldots,n}$ and
$\{\nu_l(x,t_r)\}_{l=1,\ldots,n+2}$ of $F_{n}(z,x,t_r)$ and
$H_n(z,x,t_r)$ with respect to $x$ and $t_r$ are described in terms
of the following Dubrovin-type equations.

\newtheorem{lem5.4}[lem5.1]{Lemma}
 \begin{lem5.4}
   Assume $(\ref{2.2})$ and suppose that $(\ref{5.3})$,
   $(\ref{5.4})$ hold subject to the constraint $(\ref{2.26a})$.
\begin{itemize}
  \item[\emph{(i)}]
 Suppose that the zeros $\{\mu_j(x,t_r)\}_{j=1,\ldots,n}$
 of $F_{n}(z,x,t_r)$ remain distinct for $(x,t_r) \in
 \Omega_\mu,$ where $\Omega_\mu \subseteq \mathbb{R}^2$ is
  open
 and connected, then $\{\mu_j(x,t_r)\}_{j=1,\ldots,n}$ satisfy
 the system of differential equations,
    \begin{align}
        & \mu_{j,x}=-2\frac{y(\hat{\mu}_j)}{f_0}
        \prod_{ \scriptstyle k=1 \atop \scriptstyle k \neq j}^n
        (\mu_j-\mu_k)^{-1},
        \quad j=1,\ldots,n, \label{5.56} \\
       &
       \mu_{j,t_r}=-2\frac{\widetilde{F}_{r}(\mu_j) y(\hat{\mu}_j) }
                    {f_0}
         \prod_{ \scriptstyle k=1 \atop \scriptstyle k \neq j}^n
        (\mu_j-\mu_k)^{-1},
        \quad j=1,\ldots,n,\label{5.57}
    \end{align}
with initial conditions
       \begin{equation}\label{5.58}
         \{\hat{\mu}_j(x_0,t_{0,r})\}_{j=1,\ldots,n}
         \in \mathcal{K}_{n},
       \end{equation}
for some fixed $(x_0,t_{0,r}) \in \Omega_\mu$. The initial value
problem $(\ref{5.57})$, $(\ref{5.58})$ has a unique solution
satisfying
        \begin{equation}\label{5.59}
         \hat{\mu}_j \in C^\infty(\Omega_\mu,\mathcal{K}_{n}),
         \quad j=1,\ldots,n.
        \end{equation}

   \item[\emph{(ii)}]
 Suppose that the zeros $\{\nu_l(x,t_r)\}_{l=1,\ldots,n+2}$
 of $H_n(z,x,t_r)$ remain distinct for $(x,t_r) \in
 \Omega_\nu,$ where $\Omega_\nu \subseteq \mathbb{R}^2$ is open
 and connected, then
 $\{\nu_l(x,t_r)\}_{l=1,\ldots,n+2}$ satisfy the system of
 differential equations,
     \begin{align}
        & \nu_{l,x}=2\frac{(\nu_l^2\rho^2-(u-u_{xx})\nu_l-\frac{1}{4})y(\hat{\nu}_l)}{h_0}
        \prod_{ \scriptstyle k=1 \atop \scriptstyle k \neq l}^{n+2}
        (\nu_l-\nu_k)^{-1},
        \quad l=1,\ldots,n+2, \label{5.60} \\
        &
         \nu_{l,t_r}=-2\frac{\widetilde{H}_r(\nu_l) y(\hat{\nu}_l) }
                     {h_0}
        \prod_{ \scriptstyle k=1 \atop \scriptstyle k \neq l}^{n+2}
        (\nu_l-\nu_k)^{-1},
        \quad l=1,\ldots,n+2, \label{5.61}
     \end{align}
with initial conditions
       \begin{equation}\label{5.62}
         \{\hat{\nu}_l(x_0,t_{0,r})\}_{l=1,\ldots,n+2}
         \in \mathcal{K}_{n},
       \end{equation}
for some fixed $(x_0,t_{0,r}) \in \Omega_\nu$. The initial value
problem $(\ref{5.61})$, $(\ref{5.62})$ has a unique solution
satisfying
        \begin{equation}\label{5.63}
         \hat{\nu}_l \in C^\infty(\Omega_\nu,\mathcal{K}_{n}),
         \quad l=1,\ldots,n+2.
        \end{equation}
  \end{itemize}
\end{lem5.4}
\noindent
{\it Proof.}~It suffices to prove (\ref{5.57})
 since the argument for (\ref{5.61}) is analogous
 and that for (\ref{5.56}) and (\ref{5.60})
 has been given in the proof of Lemma \ref{lemma3.4}.
 Differentiating (\ref{5.6}) with respect to $t_r$ yields
     \begin{equation}\label{5.64}
        F_{n,t_r}(\mu_j)=-f_0\mu_{j,t_r}
         \prod_{ \scriptstyle k=1 \atop \scriptstyle k \neq j}^n
        (\mu_j-\mu_k).
     \end{equation}
On the other hand, inserting $z=\mu_j$ into (\ref{5.38}) and
using (\ref{5.26}), one finds
     \begin{equation}\label{5.65}
        F_{n,t_r}(\mu_j)=2G_n(\mu_j)\widetilde{F}_{r}(\mu_j)
         =2y(\hat{\mu}_j)\widetilde{F}_{r}(\mu_j).
     \end{equation}
Combining (\ref{5.64}) and (\ref{5.65}) then yields (\ref{5.57}).
The rest is analogous to the proof of Lemma \ref{lemma3.4}.
\quad $\square$
\vspace{0.1cm}

Since the stationary trace formulas for CHD2 invariants in terms of symmetric
functions of $\mu_j$ in Lemma \ref{lemma3.5} extend line by line to the corresponding
time-dependent setting, we next record the $t_r$-dependent trace formulas without
proof. For simplicity, we confine ourselves to the simplest one only.

\newtheorem{lem5.5}[lem5.1]{Lemma}
 \begin{lem5.5}
  Assume $(\ref{2.2})$, suppose that $(\ref{5.3})$, $(\ref{5.4})$ hold, and
  let $(x,t_r) \in \mathbb{R}^2$.
  Then,
    \begin{equation}\label{5.66}
     u^{-1} \mathcal{G}(-4u_x(u-u_{xx})-2u(u_x-u_{xxx}))
   =\sum_{j=1}^n \mu_j(x,t_r)-\frac{1}{2}\sum_{m=0}^{2n+1} E_m.
    \end{equation}
 \end{lem5.5}

\section{Time-dependent algebro-geometric solutions of CHD2 hierarchy}
  In our final section, we extend the results of section 4 from the stationary
  CHD2 hierarchy,  to the time-dependent case.
  We obtain Riemann theta function representations for
  the meromorphic function $\phi$, the Baker-Akhiezer function $\psi$, and especially, for the
  algebro-geometric solutions $u, \rho$ of the whole CHD2 hierarchy.

 We first record the asymptotic properties of $\phi$ in the
  time-dependent case.

\newtheorem{lem6.1}{Lemma}[section]
   \begin{lem6.1}\label{lemma6.1}
     Assume $(\ref{2.2})$ and suppose that $(\ref{5.3})$,
     $(\ref{5.4})$ hold. Moreover, let $P=(z,y)
    \in \mathcal{K}_n \setminus \{P_{\infty_\pm},P_0\}$, $(x,t_r) \in \mathbb{R}^2$. Then,
     \begin{align}
      &
       \phi(P)\underset{\zeta \rightarrow 0}{=}
       \pm i\rho \zeta^{-1} +\frac{\mp i(u-u_{xx})-\rho_x}{2\rho}
        +O(\zeta), \quad P \rightarrow P_{\infty_\pm}, \quad \zeta=z^{-1},
        \label{6.1} \\
      &
       \phi(P)\underset{\zeta \rightarrow 0}{=}
      \frac{1}{2}+(u-u_x) \zeta+O(\zeta^2),
       \quad P \rightarrow P_0, \quad \zeta=z.\label{6.2}
     \end{align}
   \end{lem6.1}
Since the proof of Lemma \ref{lemma6.1} is identical to the corresponding stationary results
in Lemma \ref{lemma4.1}, we omit the corresponding details.

\vspace{0.1cm}

Next, we investigate the properties of the Abel map. To do this,
let $\underline{\mu}=(\mu_1,\ldots,\mu_n) \in \mathbb{C}^{n}$, we define
the following symmetric functions by
\begin{equation}\label{6.4}
        \Psi_{k}(\underline{\mu})=(-1)^{k} \sum_{\underline{l}\in \mathcal{S}_{k}}
        \mu_{l_1}\ldots \mu_{l_{k}},
        \quad
        k=1,\ldots,n,
    \end{equation}
where $\mathcal{S}_{k}=\{\underline{l}=(l_1,\ldots,l_{k}) \in \mathbb{N}^{k}
        ~|~l_1 < \ldots < l_{k} \leq n\}$;
\begin{equation}\label{6.5}
 \Phi_{k}^{(j)}(\underline{\mu})=(-1)^{k} \sum_{\underline{l}\in \mathcal{T}_{k}^{(j)}}
           \mu_{l_1}\ldots \mu_{l_{k}}, \quad
 k=1,\ldots,n-1,
  \end{equation}
where  $\mathcal{T}_{k}^{(j)}=\{\underline{l} =(l_1,\ldots,l_{k}) \in
       \mathcal{S}_{k}
      ~|~ l_m \neq j\},$ \quad $ j=1,\ldots,n.$
For the
properties of $\Psi_{k}(\underline{\mu})$ and
$\Phi_{k}^{(j)}(\underline{\mu})$, we refer to Appendix E \cite{15}.

Introducing
 \begin{equation}\label{6.7}
     \tilde{d}_{r,k}(\underline{E})
     =\sum_{s=0}^{r-k} \tilde{c}_{r-k-s} \hat{c}_s(\underline{E}),
     \quad k=0,\ldots, r\wedge n,
   \end{equation}
for a given set of constants
$ \{ \tilde{c}_l\} _{l=1,\ldots,r} \subset \mathbb{C}$,
the corresponding homogeneous and nonhomogeneous quantities
$\widehat{F}_{r}(\mu_j)$ and $ \widetilde{F}_{r}(\mu_j)$
in the CHD2 case are then given by
 \footnote{$m \wedge n = \mathrm{min}\{m,n\}$,
 $m \vee n =\mathrm{max} \{m,n\}$}
    \begin{equation}\label{6.6}
      \begin{split}
     & \widehat{F}_{r}(\mu_j)= f_0 \sum_{s=(r-n)\vee 0}^{r}
      \hat{c}_s(\underline{E})
      \Phi_{r-s}^{(j)}(\underline{\mu}),
       \\
  & \widetilde{F}_{r}(\mu_j)= \sum_{s=0}^{r} \tilde{c}_{r-s}\widehat{F}_s(\mu_j)
  =f_0
  \sum_{k=0}^{r \wedge n} \tilde{d}_{r,k}(\underline{E})
  \Phi_k^{(j)}(\underline{\mu}), \quad
    r\in \mathbb{N}_0, ~\tilde{c}_0=1,
     \end{split}
    \end{equation}
using (D.59) and (D.60) \cite{15}. Here, $\hat{c}_s(\underline{E})$, $s \in \mathbb{N}_0$,
is defined by (D.2) \cite{15}.

\vspace{0.1cm}

We now state the analog of Theorem \ref{theorem4.3}, which indicates marked differences between the CHD2
hierarchy and other completely integrable systems such as the KdV and AKNS hierarchies.

\newtheorem{the6.2}[lem6.1]{Theorem}
 \begin{the6.2}
   Assume $(\ref{2.26a})$
   and suppose that $\{\hat{\mu}_j\}_{j=1,\ldots,n}$
   satisfies the Dubrovin equations $(\ref{5.56})$, $(\ref{5.57})$
   on an open set $\Omega_\mu \subseteq \mathbb{R}^2$ such that
   $\mu_j$, $j=1,\ldots,n,$ remain distinct and nonzero on $\Omega_\mu$
   and that $\widetilde{F}_{r}(\mu_j)
   \neq 0$ on $\Omega_\mu$, $ j=1,\ldots,n$. Introducing the associated divisor
   $\mathcal{D}_{\underline{\hat{\mu}}(x,t_r)} \in \mathrm{Sym}^n(\mathcal{K}_n)$,
   one computes,
   \begin{align}
          &
             \partial_x
            \underline{\alpha}_{Q_0}(\mathcal{D}_{\underline{\hat{\mu}}(x,t_r)})
            =\frac{2}{ u(x,t_r))}
             \underline{c}(n),
            \quad    (x,t_r) \in \Omega_\mu,  \label{6.8} \\
        &
         \partial_{t_r}
         \underline{\alpha}_{Q_0}(\mathcal{D}_{  \underline{\hat{\mu}}(x,t_r)})
         =-2
          \Big( \sum_{\ell=0 \vee (n-r)}^{n}
         \tilde{d}_{r,n-\ell}(\underline{E})
         \underline{c}(\ell)
         \Big),
          \quad
         (x,t_r) \in \Omega_\mu. \label{6.9}
        \end{align}
 In particular, the Abel map dose not linearize the divisor
 $\mathcal{D}_{\underline{\hat{\mu}}(x,t_r)}$ on $\Omega_\mu$.
 \end{the6.2}
\noindent
{\it Proof.}~Let $ (x,t_r) \in \Omega_{\mu}$.
It suffices to prove (\ref{6.9}), since (\ref{6.8})
is proved as in the stationary context of Theorem
\ref{theorem4.3}.
Using  (\ref{5.57}), (\ref{6.6}), and (E.4)
\cite{15}, one infers that
   \begin{align}
    \partial_{t_r} \Big(\sum_{j=1}^n \int_{Q_0}^{\hat{\mu}_j}\underline{\omega}\Big)
    &=\sum_{j=1}^n \mu_{j,t_r} \sum_{k=1}^n
    \underline{c}(k)\frac{\mu_j^{k-1}}{y(\hat{\mu}_j)}
      \nonumber \\
    &=-2
    \sum_{j=1}^n \sum_{k=1}^n \underline{c}(k)
    \frac{\mu_j^{k-1}}{\prod_{\scriptstyle l=1 \atop \scriptstyle l \neq j}^n (\mu_j-\mu_l)}
    \frac{\widetilde{F}_{r}(\mu_j)}{f_0}
      \nonumber \\
    &=
     -\frac{2}{f_0}
     \sum_{j=1}^n \sum_{k=1}^n \underline{c}(k)
    \frac{\mu_j^{k-1}}{\prod_{\scriptstyle l=1 \atop \scriptstyle l \neq j}^n (\mu_j-\mu_l)}
     \Big(f_0
    \sum_{m=0}^{r \wedge n} \tilde{d}_{r,m}(\underline{E})
    \Phi_{m}^{(j)}(\underline{\mu}) \Big)
      \nonumber \\
       &=
     -2\sum_{m=0}^{r \wedge n} \tilde{d}_{r,m}(\underline{E})
     \sum_{k=1}^n \sum_{j=1}^n \underline{c}(k)
     (U_{n}(\underline{\mu}))_{k,j}(U_{n}(\underline{\mu}))_{j,n-m}^{-1}
       \nonumber \\
     &=
     -2\sum_{m=0}^{r \wedge n} \tilde{d}_{r,m}(\underline{E})
     \underline{c}(n-m)
        \nonumber \\
     &=
     -2\sum_{m=0\vee (n-r) }^{ n} \tilde{d}_{r,n-m}(\underline{E})
     \underline{c}(m), \label{6.12}
   \end{align}
where we used the relations (cf.(E.13), (E.14) \cite{15}),
        \begin{equation}\label{6.12a1}
         U_{n}(\underline{\mu})=
       \left( \frac{\mu_j^{k-1}}
    {\prod_{\scriptstyle l=1 \atop \scriptstyle l \neq j }^{n}(\mu_j-\mu_l)}
    \right)_{\scriptstyle j,k=1}^n, \quad
    U_{n}(\underline{\mu})^{-1}=
     \left(\Phi_{n-k}^{(j)}(\underline{\mu})\right)_{j,k=1}^n.
    \end{equation}

The analogous results hold for the corresponding divisor
$\mathcal{D}_{\underline{\hat{\nu}}(x,t_r)}$ associated with
$\phi(P,x,t_r)$.
\vspace{0.1cm}

For subsequent purpose we note the following asymptotic spectral parameter expansion of
$F_n/y$ as $P \rightarrow P_{\infty_\pm}$,
    \begin{equation}\label{6.12a2}
    \frac{F_n(z)}{y} \underset{\zeta \rightarrow 0}{=} \pm \frac{1}{2i}
    \sum_{l=0}^{\infty} \hat{f}_l \zeta^{l+1},
   \quad \textrm{as $ P \rightarrow P_{\infty_\pm}, \quad \zeta=z^{-1}. $}
    \end{equation}
Here, $\hat{f}_l$ denote the homogenous coefficients, satisfying (\ref{2.3})
with vanishing integration constants.

\vspace{0.15cm}

Next, we shall provide the explicit representations of $\psi$, $\phi$, $u,$ and $\rho$
in terms of the Riemann theta function associated with
$\mathcal{K}_n$, assuming the affine part of $\mathcal{K}_n$ to be
nonsingular.

Let $\omega_{P_{\infty_\pm}, r}^{(2)}$ be the normalized differentials of the second kind with a unique pole at
$P_{\infty_\pm}$, and principal part $\zeta^{-2-r} d\zeta$ near $P_{\infty_\pm}$, and define
    \begin{equation}\label{6.12a3}
    \widetilde{\Omega}_r^{(2)}=    \sum_{q=0}^r (q+1) \tilde{c}_{r-q}
    (\omega_{P_{\infty_-}, q}^{(2)}-\omega_{P_{\infty_+}, q}^{(2)}),
    \quad \tilde{c}_0=1,
    \end{equation}
where $\tilde{c}_q,$ $q=0,\ldots, r,$ are the constants introduced in the definition of
$\widetilde{F}_r(z)$. Hence, one infers
    \begin{align}
    & \int_{a_k}  \widetilde{\Omega}_r^{(2)}=0, \quad k=1,\ldots,n, \label{6.12a3}\\
    & \int_{Q_0}^P  \widetilde{\Omega}_r^{(2)} \underset{\zeta \rightarrow 0}{=}
      \pm \Big(\sum_{q=0}^r \tilde{c}_{r-q} \zeta^{-1-q} +\tilde{e}_{r,0}+O(\zeta)\Big),
       \quad \textrm{as $ P \rightarrow P_{\infty_\pm},$} \label{6.12a4}
    \end{align}
for some constants $\tilde{e}_{r,0} \in \mathbb{C}$.

Recalling (\ref{4.26})-(\ref{4.32}), the analog of Theorem \ref{theorem4.5} in
the stationary case then reads as follows.

\newtheorem{the6.4}[lem6.1]{Theorem}
 \begin{the6.4}
   Assume $(\ref{2.2})$ and suppose that $(\ref{5.3})$,
   $(\ref{5.4})$ hold on $\Omega$ subject to the constraint
   $(\ref{2.26a})$. In addition,
   let $P=(z,y) \in \mathcal{K}_n \setminus \{P_{\infty_\pm}\}$
   and $(x,t_r),(x_0,t_{0,r}) \in \Omega$,  where $\Omega \subseteq
   \mathbb{R}^2$ is open and connected. Moreover, suppose that
   $\mathcal{D}_{\underline{\hat{\mu}}(x,t_r)}$, or equivalently,
   $\mathcal{D}_{\underline{\hat{\nu}}(x,t_r)}$ is nonspecial for
   $(x,t_r) \in \Omega$. Then, $\phi$, $\psi$, $u$, and $\rho$ admit the
   representations
   \begin{align}
    &
    \phi(P,x,t_r)= i \rho(x,t_r)
    \frac{\theta(\underline{z}(P,\underline{\hat{\nu}}(x,t_r) ))
         \theta(\underline{z}(P_{\infty_+},\underline{\hat{\mu}}(x,t_r) ))}
     {\theta(\underline{z}(P_{\infty_+},\underline{\hat{\nu}}(x,t_r) ))
     \theta(\underline{z}(P,\underline{\hat{\mu}}(x,t_r) ))}
     \mathrm{exp} \left(d_0-\int_{Q_0}^P \Omega^{(3)}
     \right),  \label{6.19}
      \\
    &
   \psi_1(P,x,x_0,t_r,t_{0,r})=
    \frac{\theta(\underline{z}(P,\underline{\hat{\mu}}(x,t_r) ))
         \theta(\underline{z}(P_{\infty_+},\underline{\hat{\mu}}(x_0,t_{0,r}) ))}
     {\theta(\underline{z}(P_{\infty_+},\underline{\hat{\mu}}(x,t_r) ))
     \theta(\underline{z}(P,\underline{\hat{\mu}}(x_0,t_{0,r}) ))}
       \label{6.19a1} \\
    & ~~~~~~~~~~~~~~~~~~~~~~~
      \times ~
           \mathrm{exp} \left(\int_{x_0}^x dx^\prime ~i \rho(x^\prime,t_r) \int_{Q_0}^P \Omega_0^{(2)}
      +2i (t_r-t_{0,r}) \int_{Q_0}^P \widetilde{\Omega}_r^{(2)} \right), \nonumber
      \\
   &
   \psi_2(P,x,x_0,t_r,t_{0,r})=i \rho(x,t_r)
    \frac{\theta(\underline{z}(P,\underline{\hat{\nu}}(x,t_r) ))
         \theta(\underline{z}(P_{\infty_+},\underline{\hat{\mu}}(x_0,t_{0,r}) ))}
     {\theta(\underline{z}(P_{\infty_+},\underline{\hat{\nu}}(x,t_r) ))
     \theta(\underline{z}(P,\underline{\hat{\mu}}(x_0,t_{0,r}) ))}
     \mathrm{exp} \left(d_0-\int_{Q_0}^P \Omega^{(3)}
     \right)
      \nonumber  \\
    & ~~~~~~~~~~~~~~~~~~~~~~~
      \times ~
           \mathrm{exp} \left(\int_{x_0}^x dx^\prime ~i \rho(x^\prime,t_r) \int_{Q_0}^P \Omega_0^{(2)}
      +2i (t_r-t_{0,r}) \int_{Q_0}^P \widetilde{\Omega}_r^{(2)} \right),  \label{6.19a2}
     \\
   &
     u(x,t_r)=-4i
    \frac{\theta(\underline{z}(P_0,\underline{\hat{\nu}}(x,t_r) ))
         \theta(\underline{z}(P_{\infty_+},\underline{\hat{\mu}}(x,t_r) ))}
     {\theta(\underline{z}(P_{\infty_+},\underline{\hat{\nu}}(x,t_r) ))
     \theta(\underline{z}(P_0,\underline{\hat{\mu}}(x,t_r) ))},\label{6.20}
     \\
   &
    \rho(x,t_r)=-\frac{i}{2}
    \frac{\theta(\underline{z}(P_0,\underline{\hat{\mu}}(x,t_r) ))
         \theta(\underline{z}(P_{\infty_+},\underline{\hat{\nu}}(x,t_r) ))}
     {\theta(\underline{z}(P_{\infty_+},\underline{\hat{\mu}}(x,t_r) ))
     \theta(\underline{z}(P_0,\underline{\hat{\nu}}(x,t_r) ))}.\label{6.20a18}
    \end{align}
  \end{the6.4}
\noindent
{\it Proof.}~We start with the proof of the theta function representation (\ref{6.19a1}) for $\psi_1$.
Without loss of generality it suffices to treat the homogenous case
$\tilde{c}_0=1$, $\tilde{c}_q=0$, $q=1,\ldots,r$. As in the
corresponding stationary case we temporarily assume
   \begin{equation}\label{6.20a19}
   \mu_j(x, t_r) \neq \mu_{j^\prime}(x, t_r),
   \quad \textrm{for $j \neq j^\prime$ and $(x,t_r) \in \widetilde{\Omega}$}
   \end{equation}
for appropriate $ \widetilde{\Omega} \subseteq \Omega$,
and define the right-hand side of (\ref{6.19a1}) to be $\Psi$.
We intend to prove $\psi_1=\Psi$, with $\psi_1$ given by (\ref{5.22}).
For that purpose we first investigate the local zeros and poles of $\psi_1$.
Using the definition (\ref{5.24}) of $\phi$, (\ref{5.38}), and Dubrovin equations (\ref{5.57}), one computes
    \begin{align}
     &
        \widetilde{F}_r(z,x,t_r) \phi(P,x,t_r)-\widetilde{G}_r(z,x,t_r)
    =\widetilde{F}_r(z,x,t_r) \frac{y+G_n(z,x,t_r)}{F_n(z,x,t_r)}-\widetilde{G}_r(z,x,t_r)
     \nonumber \\
    &~~~~~~~~~~~~~~~~~~~
     =\frac{y ~\widetilde{F}_r(z,x,t_r) }{F_n(z,x,t_r)}+\frac{1}{2}
      \frac{F_{n,t_r}(z,x,t_r)}{F_n(z,x,t_r)}
      \nonumber \\
    &~~~~~~~~~~~~~~~~~~~
    =-\frac{1}{2} \frac{\mu_{j,t_r}}{z-\mu_j}
     -\frac{1}{2} \frac{\mu_{j,t_r}}{z-\mu_j}+O(1)
      \nonumber \\
    &~~~~~~~~~~~~~~~~~~~
    =-\frac{\mu_{j,t_r}}{z-\mu_j}+O(1),
    \quad \textrm{as $z \rightarrow \mu_j(x,t_r)$. } \label{6.21}
    \end{align}
More concisely,
    \begin{align}\label{6.22}
     \widetilde{F}_r(z,x_0,s) \phi(P,x_0,s)-\widetilde{G}_r(z,x_0,s)
     =\frac{\partial}{\partial s} \mathrm{ln} (z-\mu_j(x_0,s))+O(1)
      \nonumber \\
      \textrm{for $P$ near $\hat{\mu}_j(x_0,s)$.}
    \end{align}
Meanwhile, (\ref{4.40n2}) gives
    \begin{equation}\label{6.23}
   \phi(P,x^\prime)=\frac{\partial }{\partial x^\prime} \mathrm{ln} (z-\mu_j(x^\prime,t_r)) +O(1)
   \quad \textrm{for $P$ near $\hat{\mu}_j(x^\prime,t_r)$}.
   \end{equation}
Hence, combining (\ref{5.22}), (\ref{6.22}), and (\ref{6.23}) yields
    \begin{equation}\label{6.24}
    \psi_1(P,x,x_0,t_r,t_{0,r})=
    \left\{
      \begin{array}{ll}
        (z-\mu_j(x,t_r))O(1), & \hbox{\textrm{for $P$ near $\hat{\mu}_j(x,t_r) \neq \hat{\mu}_j(x_0,t_{0,r}),$}} \\
        O(1), & \hbox{\textrm{for $P$ near $\hat{\mu}_j(x,t_r) = \hat{\mu}_j(x_0,t_{0,r}),$}} \\
        (z-\mu_j(x_0,t_{0,r}))^{-1}O(1), & \hbox{\textrm{for $P$ near $\hat{\mu}_j(x_0,t_{0,r}) \neq \hat{\mu}_j(x,t_r),$}}
      \end{array}
    \right.
    \end{equation}
with $O(1) \neq 0$. Consequently, $\psi_1$ and $\Psi$ have identical zeros and poles on
$\mathcal{K}_n \setminus \{P_{\infty_+}, P_{\infty_-}\}$, which
are all simple by hypothesis (\ref{6.20a19}).
It remains to study the behavior of $\psi_1$ near $P_{\infty_\pm}$.
By (\ref{5.24}), (\ref{5.38}), (\ref{6.1}), and (\ref{6.12a2}), one infers that
   \begin{align}
   &
   \int_{x_0}^{x}  dx^\prime \phi(P,x^\prime,t_r)
   +\int_{t_{0,r}}^{t_r} ds ~(\widetilde{F}_r(\zeta^{-1} ,x_0,s) \phi(P,x_0,s)-\widetilde{G}_r(\zeta^{-1},x_0,s))
    \nonumber \\
   &
    \underset{\zeta \rightarrow 0}{=}
    \pm i \zeta^{-1} \int_{x_0}^{x} \rho(x^\prime,t_r)~ dx^\prime
    +\left\{
       \begin{array}{ll}
         O(1) & \hbox{\textrm{for $P \rightarrow P_{\infty_+}$ }} \\
         O(1) & \hbox{\textrm{for $P \rightarrow P_{\infty_-}$ }}
       \end{array}
     \right.
     \nonumber \\
   &~~~~~~
    +\int_{t_{0,r}}^{t_r} ds \left(
    \frac{y~\widetilde{F}_r(\zeta^{-1} ,x_0,s)}{F_n(\zeta^{-1} ,x_0,s)}
    +\frac{1}{2}\frac{F_{n,t_r}(\zeta^{-1} ,x_0,s)}{F_n(\zeta^{-1} ,x_0,s)}
     \right)
     \nonumber \\
  &
    \underset{\zeta \rightarrow 0}{=}
    \pm i \zeta^{-1} \int_{x_0}^{x} \rho(x^\prime,t_r)~ dx^\prime
    +\left\{
       \begin{array}{ll}
         O(1) & \hbox{\textrm{for $P \rightarrow P_{\infty_+}$ }} \\
         O(1) & \hbox{\textrm{for $P \rightarrow P_{\infty_-}$ }}
       \end{array}
     \right.
     \nonumber \\
  &~~~~~~
    +\int_{t_{0,r}}^{t_r} ds \left( \pm 2i \zeta^{-r-1}
    \frac{\sum_{m=0}^r \tilde{f}_m (x_0,s) \zeta^m}{\sum_{l=0}^\infty \tilde{f}_l (x_0,s) \zeta^l}
    +\frac{1}{2} \frac{u_{t_r}(x_0,s)}{u(x_0,s)}+O(\zeta)
    \right)
     \nonumber \\
  &
    \underset{\zeta \rightarrow 0}{=}
    \pm i \zeta^{-1} \int_{x_0}^{x} \rho(x^\prime,t_r)~ dx^\prime
    +\left\{
       \begin{array}{ll}
         O(1) & \hbox{\textrm{for $P \rightarrow P_{\infty_+}$ }} \\
         O(1) & \hbox{\textrm{for $P \rightarrow P_{\infty_-}$ }}
       \end{array}
     \right.
     \nonumber \\
    &~~~~~~
    +\int_{t_{0,r}}^{t_r} ds \left( \pm 2i \zeta^{-r-1}
    \mp \frac{2i \tilde{f}_{r+1}(x_0,s)}{\tilde{f}_0(x_0,s)}
    +\frac{1}{2} \frac{u_{t_r}(x_0,s)}{u(x_0,s)}+O(\zeta)
    \right)
    \nonumber \\
    &
    \underset{\zeta \rightarrow 0}{=}
    \pm i \zeta^{-1} \int_{x_0}^{x} \rho(x^\prime,t_r)~ dx^\prime
    \pm 2i \zeta^{-r-1} (t_r-t_{0,r})
    +\left\{
       \begin{array}{ll}
         O(1) & \hbox{\textrm{for $P \rightarrow P_{\infty_+}$, }} \\
         O(1) & \hbox{\textrm{for $P \rightarrow P_{\infty_-}$, }}
       \end{array}
          \right. \label{6.25}
   \end{align}
where we used $\tilde{f}_0=-u$ and $u_{t_r}=\frac{1}{2}\tilde{f}_{r,x}$ (cf.(\ref{5.1}))
in the homogeneous case $\tilde{c}_0=1$, $\tilde{c}_q=0$, $q=1,\ldots,r$.
A comparison of $\psi_1$ and $\Psi$ near $P_{\infty_\pm}$,
taking into account (\ref{5.22}), (\ref{6.12a4}), (\ref{6.19a1}), and (\ref{6.25}),
then shows that $\psi_1$ and $\Psi$ have identical exponential behavior up to order $O(1)$
near $P_{\infty_\pm}$. Thus, $\psi_1$ and $\Psi$ share
the same singularities and zeros, and the Riemann-Roch-type uniqueness result
(cf. Lemma C.2 \cite{15}) then proves $\psi_1=\Psi$.
The representation (\ref{6.19}) for $\phi$ on $\widetilde{\Omega}$
follows by combining (\ref{5.28}), (\ref{6.1}), and
Theorem A.26 \cite{15}. The representation (\ref{6.20a18}) for $\rho$ on
$\widetilde{\Omega}$ is clear from (\ref{6.19}) and (\ref{6.2}).
The representation (\ref{6.20}) for $u$ on
$\widetilde{\Omega}$ follows from (\ref{6.20a18}) and the relation
$u\rho=-2$.
In fact, since the proofs of (\ref{6.19}), (\ref{6.20}), and (\ref{6.20a18})
are identical to the corresponding stationary results
in Theorem \ref{theorem4.5}, which can be extended line by line to the
time-dependent setting, here we omit the corresponding details.
Finally, the extension of all these results from
$(x,t_r) \in \widetilde{\Omega}$ to $(x,t_r) \in \Omega$ then simply follows from
the continuity of $\underline{\alpha}_{Q_0}$ and the hypothesis of
$\mathcal{D}_{\hat{\mu}(x,t_r)}$ being nonspecial for  $(x,t_r) \in \Omega$. \quad $\square$

\vspace{0.15cm}

 Remark $\ref{remark4.8}$ applies in the  present time-dependent
    context as well.

\section*{Acknowledgments}
The work described in this paper
was supported by grants from the National Science
Foundation of China (Project No.10971031; 11271079; 11075055),
Doctoral Programs Foundation of the Ministry of Education of China,
and the Shanghai Shuguang
Tracking Project (Project 08GG01).

\end{document}